\documentstyle[sprocl,psfig]{article}
     \newcommand{\ds}{\displaystyle}
     
\begin{document}\hbadness=10000
\title{RELATIVISTIC QUARK PHYSICS
}

\author{Johann Rafelski
}

\address{Department of Physics, University of Arizona, 
        Tucson, AZ, 85721}

$ $\\
\begin{center}\vskip -5cm{\bf Presented in August 1997 at
the 5th Rio de Janeiro International\\ Workshop on\\
Relativistic Aspects of Nuclear Physics}\vskip 2.7cm
\end{center}
\maketitle\abstracts{
We present a brief  survey of the development of 
nuclear physics towards relativistic quark physics. This 
is followed by   a   
thorough discussion of the quest for the observation of the
dissolution of nuclear matter into the deconfined quark matter 
(QGP) in relativistic nuclear  collisions. Use of strange 
particle signatures in search for QGP is emphasized.
}

\section{From Nuclear to Quark Physics}\label{intro}
\subsubsection{What (we think) we know and what we know we do not know}
The atomic nucleus is a quantum bound state of nucleons 
(protons and neutrons) each comprising three confined i.e. 
permanently bound valance quarks. 
The dynamics of nearly massless quarks $q=u,\,d$ inside
a nucleon is fully relativistic. Thus the mass of the nucleon 
and hence of the nucleus, and therefore of about 98\% of all 
matter known to us, is understood to be the relativistic quark
quantum-zero-point energy. Confinement localizes quarks to within
a volume with radius of about 1\,fm in size, about 20--40 times 
less than  (light $q=u,\,d$) 
quark Compton wavelength $\lambda(q)=\hbar c/m_q$, with 
$3<m_q<15$\,MeV.  Most important point here is that the 
indirectly arising interactions, due to the properties of 
the confining strongly interacting vacuum state,
determine the scale of hadronic size, and thus mass, and
thus all the properties of the Universe we see around us.
Moreover, as we shall argue at the end, there is all reason
to believe that the hadronic energy scale is a direct descendent
the unification scale of gauge interactions.

Research of past 50 years, since the discovery of the 
pion~\cite{pidisc}, established the following  paradigm of the 
properties of strong interactions~\cite{FGL73,Wei96}:
 SU(3)$_{\rm c}$-QCD
is the fundamental gauge theory of strong interactions, with 
quarks $q$ and gluons $G$ being the {\it color} charged
fundamental degrees of freedom; all strongly interacting 
and in our vacuum mobile particles are color neutral.
To explain the confinement of colored states to localized
region of space time the principal postulate is~\cite{Lee82} that 
the {\it true} vacuum  state $|V\rangle$ abhors the color charge; 
but there is an excited  state $|P\rangle$ referred to as {\it
perturbative vacuum} in which colored $q,\,G$ particles 
are mobile and have their naive perturbative physical relevance.

The physical difference between $|P\rangle$ and $|V\rangle$
is akin to the difference between vapor and ice;
in regions of space-time in which `elementary' vapor exists,
where $|P\rangle$ replaces $|V\rangle$,  a finite energy density 
the latent heat $\cal B$ has been deposited.
Pushing  the vapor and liquid phase similarity 
we see that there may be an intermediate state of the vacuum
which we shall call $|F\rangle$. In this 
`fluid flavor state' the color quantum numbers
of quark and gluons are mobile, but the dissolution of $|V\rangle$ 
is not complete, and relatively little latent heat is
consumed in the transformation $|V\rangle\to|F\rangle$. The principal
degrees of freedom in the $|F\rangle$-vacuum could be effective
`constituent' quarks. 

Thus to dissolve fully the quantum vacuum structure a considerable latent 
(condensation) heat per unit volume  needs to be delivered.
It is thought that it is of the magnitude $\cal B$=0.1--1~GeV\,fm$^{-3}$. The
lower value ${\cal B}_{\rm F}=0.1$ GeV\,fm$^{-3}$ is the energy scale
required for the understanding of
hadron masses (see Section \ref{BagM}), the upper value
${\cal B}_{\rm P} =1$~GeV\,fm$^{-3}$  
arises aside from lattice gauge studies of QCD~\cite{lattice}
from the necessity to make the deconfining, perturbative state $|P\rangle$
inaccessible in low energy reactions, in which free quarks
and gluons do not become apparent as the fundamental degrees of
freedom. ${\cal B}_{\rm F}$  is the dimensioned quantity which
determines the hadronic mass and size scales. The value of
${\cal B}_{\rm F}=0.1$ GeV\,fm$^{-3}$ 
has to be compared to `normal' matter condensation energies which have a
magnitude of say 0.1eV \AA$^{-3}$, a factor $10^{24}$ smaller!
In order to study states with extreme properties rivaling those
found in early Universe some 40\,$\mu$s after its birth
we have developed experimental facilities in which heaviest
nuclei are made to  collide at high energies. 
 
\begin{table*}[!t]
\caption{Strong interaction vacuum states}
\vspace{0.4cm}
\begin{center} 
\begin{tabular}{|c||c|c|c|} 
\hline\vphantom{$\displaystyle\frac{E}{B}$}
vacuum&state& $E/V$~[GeV\,fm$^{-3}$] & $m_{\rm q}$~[MeV] \\
\hline\hline\vphantom{$\displaystyle\frac{E}{B}$}
true & $|V\rangle$ & 0 & $\infty$\\
\hline\vphantom{$\displaystyle\frac{E}{B}$}
fluid flavor & $|F\rangle$ & ${\cal B}_{\rm F}=0.1$ & 350\\
\hline\vphantom{$\displaystyle\frac{E}{B}$}
perturbative & $|P\rangle$ & ${\cal B}_{\rm P} =1 $ & 5--15\\
\hline
\end{tabular} 
\end{center} 
\end{table*} 

\subsubsection{Unclear horizons}
There are many open questions and unresolved puzzles
in relativistic quark physics. Some are fundamental and 
border on the comprehension of the origin of the standard model,
other are simply consequences of the complexity of the gauge 
theory that governs quantum-chromodynamics (QCD), the theory
of strong interactions. 
\begin{itemize}
\item The fundamental questions are
of the type: why do quarks have so many different quantum numbers,
 e.g. i) fractional electrical charge, ii) color non-Abelian charge, 
iii) several (three) families of particles,
 where $(u,d)$, $(c,s)$ and $(t,b)$ 
are the three doublets of quarks, iv)  intrinsic spin. 
However, an elementary object is expected to posses just a few simple 
properties. Are perhaps even quarks not elementary?
\item
Understanding of quark confinement 
riddle falls under the practical category of solving QCD. 
It is not merely the problem of the vacuum properties, but 
significantly, a problem of hadronic spectrum. The key problem that
emerges is that we cannot find experimentally many of the bound
states of quarks that intuitive extrapolations of the current 
understanding imply. 
\end{itemize}
Diverse simple quark bound states other than known
mesons and baryons corresponding to angular and 
radial quark excitations inside 
the confining boundary are absent. Other  expected
states, such as the nucleon-anti-nucleon molecule (baryonium)
which in the quark language consists of the spatially separate
two quark -- two antiquark clusters, $qq$--$\bar q\bar q$, 
have also not been seen. 

The nucleus is made of $3N$ quarks, but quarks remain  clustered 
in nucleons. Why are quarks  not seen to bind other than in clusters of
three? How much excited would be the state comprising $3N$ freely 
movable quarks? In a two-vacuum model this is a state that should
be nearly stable. But it is hard to imagine formation of dense quark matter 
without allowing for heating during compression in collision
of heavy nuclei. At finite temperature we
have some quark-antiquark pairs and real gluons appear. 
This is the quark-gluon plasma (QGP) state of matter last seen in 
the early Universe. In the limit that 
color charge interactions of quarks and gluons are small, 
which we believe to be the case in non-Abelian gauge theories 
at high energies, given the decrease in strength with energy scale of 
the coupling constant, the
relativistic quantum gas properties only depend on the number of 
excitable degrees of freedom, that is the degeneracy $g$
and the kinematic relation between energy 
and particle momentum. 

It is of course far from certain that the QGP can or/and will be 
created in relativistic nuclear collisions which last just an instant, 
as short as light needs to cross a nucleus, 
$10^{-22}$\,s. Our hope and expectation is that in a 
statistical system with many degrees of freedom at as high a temperature 
as $kT=150$--$250$ MeV, the detail of hadron dynamics and structure 
that escaped our attention will become irrelevant and we will be able
to observe properties of deconfined nuclear matter. It is possible
that the complex aspects of
interactions within  confined bound states become irrelevant in the 
high density/energy limit. The question then is under what conditions 
we actually encounter this asymptotic limit, and what, if anything, 
we can learn  when studying the approach to this limiting case.

We shall address here several issues that we left open in our report 
made at the 
last meeting~\cite{Rio95} and we shall attempt to make an elementary 
presentation accessible in a large part to students; however, we
recommend that the second half  of this paper be read in conjunction
with our earlier report~\cite{Rio95}, as it continues and updates 
the developments reported. Before proceeding with
more rigorous discussion of the vacuum and hadronic structure in 
Section \ref{BagM}, we shall give next a brief historical reminiscence
about relativistic nuclear physics. Recent aspects of quark-gluon plasma 
studies are introduced in Section \ref{QGPsec}. We describe the 
considerable progress that has occured in past tow years regarding
strange particle flavor production in Section \ref{sprod}. The still
preliminary discussion of particle production in hadronizing  QGP
is offered in Section \ref{hadro}, where we also compare with recent
Pb--Pb 158 GeV A CERN heavy ion collision experimental results. 
We conclude this report with a discussion of the  ongoing research
work.

\subsubsection{Historical remarks}
The discovery of natural radioactivity more than 
100 years ago occurred when both quantum 
mechanics and relativity, the two pillars of our current quest
for quark physics, were not yet formulated in their final form. 
Relativity and nuclear physics evolved  initially
without mutual interaction. Quantum physics was applied rapidly
to understand puzzles of nuclear physics, a good example is
the $\alpha$-decay using quantum tunneling. Relativity 
enters with  Fermi theory of  $\beta$-decay: {\it `...in order
to obtain relativistically invariant form ... necessary at the
velocities of emitted  electrons close to velocity of light...
we must use Dirac four functions...'} writes George Gamov
in {\it Structure of Atomic Nuclei and Nuclear Transformations}
published 60 years ago. This is  perhaps the first 
textbook mention of  relativity within nuclear physics.

Hideki Yukawa proposed a theoretical yet undiscovered
particle, the meson, as the origin of short range 
nuclear interactions. This bold step generated 
a lot of interest in fundamental understanding of 
nuclear forces and nuclei, which continues to this day.
After the initial confusion caused by the unexpected 
heavy electron, the muon, the 
discovery of the $\pi$-meson in 1947~\cite{pidisc} 
by Lattes, Occhialini and Powell  just 
50 years ago today made the odd couple, 
relativity and nuclear physics, inseparable. With this discovery 
a novel interpretation of the nucleon-nucleon interaction came
immediately within range.  Already in 1948,  Rosenfeld's
 book {\it Nuclear Forces} describes
the Breit-type  reduction of the fundamental meson interaction to 
the effective nonrelativistic form, involving the spin-orbit coupling.

Since there are several mesons, generalization of Yukawa ideas
involve nucleon-nucleon interactions with different Lorentz
transformation symmetry, from which the nuclear shell model 
potential arises, given that the range of the interaction is 
the mean distance between nucleons in nuclei. In magnitude, 
the effective single particle nucleon potential $V_{\rm eff}$
in the nucleus is not much bigger than a few percent 
of nucleon mass $m_{\rm N}$, as is born out by Bethe-Weizs\"acker 
mass formula which gives the bulk nuclear binding energy at
about 15 MeV/nucleon, less than 2\% of $m_{\rm N}$. 
However, the spin-orbit coupling influence on nucleon energies 
is in comparison surprisingly big, as is born out by study of 
nuclear spectra. Relativity provides here a rather simple 
explanation of a fundamental puzzle of nuclear physics, and
also explains how it can happen that strongly interacting 
nucleons are bound by a relatively small potential.
At this point relativity becomes inseparable part of 
nuclear physics.

The solution of this puzzle 
requires consistent Breit reduction of relativistic 
wave mechanics to non-relativistic limit, schematically:
\begin{equation}
\label{Veff}
V_{\rm eff}\simeq \{-U+V\}+\frac{1}{2m_{\rm N}^2r}\frac{d}{dr}
                \{ U+V\}\vec L\cdot \vec S +\ldots\,.
\end{equation}
We see that the  spin-orbit force is
the sum of the gradients of the (pseudo)scalar $U$ and 
vector meson $V$ exchange potentials, while the radial 
part of the effective 
potential for the nucleons is the difference of both 
these contributions $V-U$. This than explains the relatively 
small nuclear shell potential, now understood as
the difference $V-U$, as $U,V$ can be large as long as their
magnitude is similar. Indeed, we can have complete absence of 
spin-orbit coupling \protect~\cite{Sof73} when the $V,U$ potentials add and 
thus the $\vec L\cdot \vec S$-coupling term vanishes, while the opposite 
limit in which the (nearly) exact  cancelation of the $V,U$ potentials, 
and the associated maximum of the $\vec L\cdot \vec S$ occurs, is the
interesting property of nuclear interactions. Today, we believe that
these nuclear interactions
are the  `unscreened', van der Waals-type forces between quarks
confined to individual nucleons. Efforts continue to obtain
QCD-quark physics based derivation of the  residual nuclear
interactions. 

The development of nuclear matter theory, neutron matter in neutron stars
ensued in parallel with studies of high energy cosmic ray reactions.
These two initially separate developments were precursors to 
our present day interests in hot nuclear matter, and ultimately hot 
hadronic matter, that is nuclear matter at so high temperature that it
contains a significant meson abundance. On the other side, there was
early recognition among particle physicists that elementary collisions
involving strongly interacting particles lead to multi-particle production
which could be described as if originating from a fireball of dense,
hot matter. Such  theoretical work on multi-particle
production by E. Fermi~\cite{Fer50} in USA, and L. Landau~\cite{Lan53}
in Russia, which paved the way  to the development in early
sixties of statistical bootstrap model description of hadron production.
Rolf Hagedorn, working out of CERN, has  been since 1964
implementing a  theoretical picture of particle production
from boiling hot hadronic matter~\cite{HAG}.  He was able to describe
many experimental features of hadronic spectra within the newly developed 
statistical bootstrap model of dense hadronic matter comprising 
a resonance gas of point-like particles. Not only the
properties of hadronic matter but also the hadronic mass spectrum 
emerged correctly from this approach. Ultimately, a connection 
has been established between particle and nuclear 
hot hadronic matter descriptions, by way of introducing into Hagedorn's
bootstrap aside of 'elementary' mesons, also 
'elementary' baryons of finite size~\cite{HR80}.

In parallel to the development of phenomenological hadronic matter 
theory the fundamental understanding of hadronic structure was 
emerging. Just more than 25 years ago the present day stage was set when
quantum-chromo\-dynamics (QCD) was recognized as the fundamental 
gauge theory of strong interactions~\cite{FGL73} from 
a number of other attempts to explain hadronic structure
in terms of quarks. The development of hadronic structure 
within the MIT-bag model followed without delay~\cite{Cho74},
with constituents of  nucleons, $u$p and $d$own quarks having in
this approach just very
tiny masses.  To explain why the nucleon is so much
more heavy than an electron an old model of the neutron has been 
resurrected: before
the neutron discovery, the atomic nucleus was constructed from
 protons and tightly bound additional electrons. Without quarks 
this model could 
not work well since electrons were shown not to have strong 
interactions. Replacing the electrons with quarks which have indeed
both strong and electric charge allows to describe the large masses 
of protons and neutrons in terms of the light elementary object, 
the `electron' of QCD,  the quark. 

With light quarks as building blocks of hadrons, it took no time
at all for Carruthers to propose the existence of 
``Quarkium: a bizzare Fermi liquid''~\cite{Car73}. 
Hagedorn extended his statistical bootstrap
 by considering the interior of hadrons, which
were now understood as having a finite volume filled with light quarks. 
Under Hagedorn's guidance, I begun in 1977 at CERN to 
explore the  melting/dissolution/fusion of dense and hot hadronic 
bootstrap matter into quark matter.
 One of the key issues is, if at the  conditions of 
density and temperature, when confined, nucleon/meson type matter, 
reach the crossover point to  the color conductive 
conditions of the vacuum we actually will encounter 
a phase transition.  We soon arrived 
at a detailed description of the first order 
phase transition that we 
believed occurs between the confined 
and deconfined phase~\cite{HR81}. Though this subject has since 
many times been revisited, this extension
of the statistical bootstrap model to finite size
hadrons with quark internal structure is still
today the foundation of all detailed
models of the dense hadronic matter 
formed in relativistic nuclear  collision. 

\subsubsection{First conclusions: relativistic nuclear collisions}
Looking back to these pioneering times of quark matter 
days  the text of  the abstract of the  inaugural lecture 
of June 1980, which I presented at Frankfurt University
comes to mind. It is quite short:  
{\it Quark Matter -- Nuclear Matter}: 
`The fusion of  the  constituents -- quarks --- 
 of protons and neutrons
into quark matter, a new phase of 
nuclear matter, is expected 
to occur under experimentally 
accessible conditions of pressure and
temperature.' 

Our objective is to\\
-- break up matter, freeing quarks,
in laboratory experiments;\\
-- study 
the Universe about 40$\mu$s after the Big-Bang by measuring 
properties of this hot, deconfined state;\\
-- create a firm foundation of understanding of strong interactions
in terms of a fundamental theoretical paradigm, confirmed 
experimentally.\\ 
Nearly certainly, there is some not yet 
understood principle at work that is needed to eliminate unobserved
hadronic states. This is at the same time a challenge and 
an undesirable complication within this research program.

\section{Quark Bag Model}\label{BagM}
\subsubsection{Vacuum Structure}
Among the most far-reaching developments of the recent 25 years of 
research into the consequences of fundamental interactions is 
recognition that the true physical vacuum is a state of considerable 
complex and physically significant structure. While we  know that
strictly speaking  the vacuum is empty, its quantum structure 
(wave function) can be highly non-trivial, deviating considerably  
from that of a non-interacting Fock space. This is not the place to 
present the multitude of phenomena that go along with this 
effect, but we can illustrate some simple properties in order to 
justify the principal of confinement and the related appearance of 
light relativistic quarks in the description of the deconfined QGP. 
Indeed, one finds a quite remarkable comment to this point in 
Weinberg treatise~\cite{Wei96} (Volume II, p190, bottom) on Quantum Theory
of Fields: `{\ldots this work was done without a specific theory of the 
strong interactions. One of the reasons for the rapid acceptance 
of quantum chromodynamics in 1973 as the correct theory of strong 
interactions was that it explained the SU(2)xSU(2) symmetry 
[inherent in Adler-Weisberger sum rule of 1965] as a simple consequence
of the smallness of the $u$ and $d$ quark masses.}' So if quarks have a 
small mass, what is the origin of the 1 GeV scale of the mass of nucleons? 

The infrared QCD instability leads to the appearance of a 
finite glue `condensate'  
field i.e vacuum expectation value (VEV) of the gluon field-correlator
~\cite{SVZ79,Shif92}, which is evaluated from experimental data to be:
\begin{equation}\label{Gcon}
\langle \frac{\alpha_s}{\pi} G^2\rangle\simeq (2.3\pm0.3) 10^{-2}{\rm GeV}^4
=[390\pm12\,{\rm MeV}]^4\,,
\end{equation}
here $\alpha_s=g^2/4\pi$ is the strong interaction (running) 
coupling constant, see Section \ref{sprod}, and 
\begin{eqnarray}\label{G2}
\frac12 G^2\equiv \frac12 \sum_aG^a_{\mu\nu}G_a^{\mu\nu}
   =\sum_a[\vec B_a^{\,2}-\vec E_a^{\,2}]\,,
\end{eqnarray}
with $a=1, \ldots, N^2-1)$ gauge field components in SU$_c(N)$. 
The value (and space-time shape) of the  glue condensate has also been 
extracted from lattice gauge calculations~\cite{EdGM98}. 
It agrees well with the numerical value shown in Eq.\,(\ref{Gcon}),
obtained from QCD sum-rules~\cite{SVZ79,Nar96}. Note that 
Lorentz and gauge symmetry of the vacuum state dictates that the VEV of a
product of two  field operators  of the type shown here satisfy:
\begin{equation}\label{EBconG}
\langle  G^a_{\mu\nu}(x) G^b_{\rho\sigma}(x)\rangle=
(g_{\mu\rho}g_{\nu\sigma}-g_{\mu\sigma} g_{\nu\rho})\delta^{ab}
\frac{1}{96} \langle G^2(x)\rangle\,,
\end{equation}
Taking the required contractions and using Eq.\,(\ref{G2}) one finds:
\begin{equation}\label{EBcon}
\langle \sum_a\vec B_a^{\,2}\rangle=-\langle \sum_a\vec E_a^{\,2}\rangle
\end{equation}
Eq.\,(\ref{Gcon}) implies that $\langle \vec B^{\, 2}_a(x)\rangle$ 
is positive, and $\langle\vec E^{\,2}_a(x)\rangle$ is negative. 
However, these signs are defined with
reference to the perturbative state $|P\rangle$. Specifically, we are 
considering products of field operators which are normal-ordered with 
respect to $|P\rangle$ -- where the condensate fields thus 
vanish by definition. 
The interpretation  of Eq.\,(\ref{EBcon}) is that the B-field
fluctuates in the true QCD vacuum $|V\rangle$ 
with a bigger amplitude than in the 
perturbative `vacuum' $|P\rangle$, while the E-field fluctuates with a
smaller amplitude than in $|P\rangle$: $|V\rangle$, the true 
vacuum, may be completely  `magnetic', without any
electric fluctuations, while $|P\rangle$ may be 
completely` electric', without any magnetic fluctuations. 
There are obviously many different  ways to model and understand the 
glue condensate,  and we shall not discuss this here in any greater detail.
Within such models a clear connection between the condensation energy $\cal B$ and
the glue condensate $\langle \frac{\alpha_s}{\pi} G^2\rangle$ usually arises.

Another important property of the true vacuum has been shown well
before the fundamental degrees of freedom and QCD  have been understood.
Expressed in modern language, the GOR-current algebra relation reads
(see e.g~\cite{Kle92}): 
\begin{equation}\label{udconGOR}
m_\pi^2f_\pi^2\simeq-\frac12(m_u+m_d)\langle \bar uu+\bar dd\rangle\,,
\end{equation}
where $u\,,d$ are 
the spinor  field  operators representing the two light quark
flavor fields.  
The pion decay constant is $f_\pi=93.3$\,MeV, and $m_\pi$ refers 
to the charged pion mass $m_{\pi^+}=139.6$\,MeV. The non-vanishing light
quark masses $m_u+m_d$ are recognized to be the source of a 
finite mass of the pion; 
for vanishing quark masses the hadronic chiral SU(2)$_L$xSU(2)$_R$ 
symmetry is realized in QCD is exactly.
Current best estimate of the running QCD-current quark masses at 
1\,GeV scale are~\cite{Nar96,DN98,FK98}:
\begin{equation}\label{udmass}
(m_u+m_d)\vert_{1\,{\rm GeV}}\simeq 14.7\pm0.8\,\mbox{MeV}\,,\qquad
m_s\vert_{1\,{\rm GeV}}\simeq 195\pm12\,\mbox{MeV}\,.
\end{equation}
This light quark mass estimate arises from the 
GOR relation Eq.\,(\ref{udconGOR})
and the quark condensate deduced from the properties of the vacuum: 
\begin{equation}\label{udcon}
\frac12\langle \bar uu+\bar dd\rangle\vert_{1\,{\rm GeV}}
  \equiv \frac12\langle \bar q q\rangle\vert_{1\,{\rm GeV}}
   =-[(225\pm9)\,{\rm MeV}]^3\,.
\end{equation}

Model calculations~\cite{DS81,Sot85,EN98} employing mean field
configurations of gauge fields in the QCD vacuum invariably 
suggest that it is the presence of the glue field condensate which
is the driving force behind  the appearance of the 
quark condensate. For example a specific  model~\cite{ES86}, 
(which employed a self-dual covariantly constant field)  for the 
non-perturbative gauge field configurations in the 
structured QCD vacuum, finds that the quark condensation is a minor
and stabilizing contributor (6\%) to the vacuum energy due in  its bulk 
part to the glue degrees of freedom. We further note that at high 
temperature the vacuum structure of QCD, 
as expressed by Eq.\,(\ref{Gcon}) in terms of the glue condensate, 
melts and one reaches the perturbative vacuum~\cite{lattice}.
This confinement to deconfinement transformation and the chiral 
symmetry restoration as expressed by the melting of the quark condensate 
are seen at the same temperature.

It is thus generally assumed that the glue condensate is driving
the quark condensate, and thus the large chiral symmetry breaking,
despite the smallness of the quark masses.  This is tantamount to 
the picture of quarks confined by glue fluctuations and the resulting
large relativistic  quark confinement energy. We now illustrate 
how presence of (fluctuating)  gauge fields can induce appearance of
Fermion condensates. Schwinger in his seminal paper on gauge invariance 
and vacuum fluctuations [see~\cite{Sch51}, Eq.\,(5.2)] shows:
\begin{equation}\label{Psicon}
\langle\bar \psi(x)\psi(x)\rangle\to\frac12\langle[\bar \psi(x),\psi(x)]\rangle
= -{{\partial \Gamma[A_\mu]}\over {\partial m}}
\end{equation}
The left-hand-side of Eq.\,(\ref{Psicon}) 
defines here more precisely the meaning of the quark 
condensate in terms of the  Fermi field operators at equal space-time point.
The right-hand-side refers to the effective action density $\Gamma[A_\mu]$ 
of Fermions in presence of gauge potentials $A_\mu$.

The one loop the effective action $\Gamma^{(1)}$ for 
constant gauge fields in the  Abelian (QED) theory was already 
presented by Euler, Heisenberg, Kockel~\cite{HE36}. We re-express 
this result using the invariant Maxwell-field-like-quantities 
$E,B$, which are related to the 
Maxwellian electric $\vec E$ and magnetic $\vec B$ fields by: 
\begin{eqnarray}\label{conB}
B^2&=&\frac{e^2}2\sqrt{(\vec E^{\,2}-\vec B^{\,2})^2+4(\vec E\cdot \vec B)^2}-
\frac{e^2}2(\vec E^{\,2}-\vec B^{\,2})\\
\nonumber &\to&|e\vec B|^2\,,\qquad 
  {\rm for}\ |\vec E|\to 0\,;\\
\phantom{65}\nonumber\\
E^2&=&\frac{e^2}2\sqrt{(\vec E^{\,2}-\vec B^{\,2})^2+4(\vec E\cdot \vec B)^2}+
\frac{e^2}2(\vec E^{\,2}-\vec B^{\,2})\\
\nonumber &\to& |e\vec E|^2\,,\qquad 
  {\rm for}\ |\vec B|\to 0\,.
\label{conE}\end{eqnarray}
We also recall for later convenience:
\begin{eqnarray}\label{EF2}
\vec E^{\,2}-\vec B^{\,2} = -\frac12 F_{\mu\nu}F^{\mu\nu}
   \equiv -\frac12 F^2\,,\qquad
\vec E\cdot \vec B=-\frac14  F_{\mu\nu}\tilde{F}^{\mu\nu}
    \equiv -\frac14 F\tilde{F}\,,
\end{eqnarray}
where:
\begin{equation}\label{Fdef}
F^{\mu\nu}=\partial^\mu A^\nu-\partial^\nu A^\mu\,,\qquad
\tilde{F}^{\mu\nu}=\frac12\epsilon^{\mu\nu\alpha\beta} F_{\alpha\beta}
\end{equation}

In this notation the effective QED action to first order in 
the fine-structure constant $\alpha_e=e^2/4\pi$, 
and evaluated in the limit that
the Maxwell field is constant on the scale of electron's Compton 
wave length (which is the situation for all externally applied 
macroscopic fields) is given by:
\begin{equation}\label{EHfull}
\Gamma^{(1)}=-\frac1{8\pi^2}\int_0^\infty\frac{ds}{s^3}{\rm e}^{-m^2s}
  \left[\frac{sE}{\tan sE}\,\frac{sB}{\tanh sB}
  -1+\frac13(E^2-B^2)s^2\right]\,.
\end{equation}   
We note the two-fold subtraction required here: the first eliminates the 
field independent, zero-point action of the perturbative vacuum, and 
corresponds to normal-ordering of the field operators. The second 
subtraction is absorbed in  charge
renormalization which assures that for weak fields the  perturbative 
asymptotic series begins ${\cal O}(E^4,B^4,E^2B^2)$. Schwinger~\cite{Sch51}
pointed out that the singularities along the real $s$-axis of the 
proper-time integral in Eq.\,(\ref{EHfull}) are related to instability 
of the vacuum due to pair production, a process in principle 
possible when potentials are present~\cite{RFK78} that can rise 
more than $2m$, which is of course the case in presence of a constant, 
infinite range, electrical fields.

Employing  Eq.\,(\ref{Psicon}) we obtain the  condensate:
\begin{eqnarray}\label{PsiconEB}
-m\frac12\langle[\bar \psi(x),\psi(x)]\rangle^{(1)}=
  \frac{m^2}{4\pi^2}\int_0^\infty\frac {ds}{s^2} {\rm e}^{-m^2s}
   \left[\frac{sE}{\tan sE}\,\frac{sB}{\tanh sB} -1
\right]\,. \end{eqnarray}
We omitted here the renormalization of charge subtraction  term 
[second subtraction in  Eq.\,(\ref{PsiconEB})] 
needed only for charge renormalization.
Schwinger~\cite{Sch51} reintroduced this specific term to obtain the leading 
 effective  electromagnetic interaction coupling  of a neutral scalar 
meson to two photons.

There are few if any  observable macroscopic effects of the fermion
vacuum fluctuation in QED as these need to be induced by an extremely strong 
external force. 
In QCD situation is fundamentally different, due to spontaneous 
field fluctuations in the glue sector. To obtain a measure of
the effects we need to substitute in the
effective action Eq.\,(\ref{EHfull}) the pure U(1) Maxwell gauge field by 
the invariant combination of U(1) and the condensed SU(3) gauge fields:
\begin{equation} \label{twofields}
 \frac{\alpha}{\pi}F^2\to 
 \langle\frac{\alpha_s}{\pi} G^2\rangle+ \frac{\alpha_e}{\pi} q^2F^2\,.
\end{equation}
Here $q=2/3,\,-1/3$ are the  quark fractional Maxwell charges.
In general one can `forget' about the QED charges and 
proceed with QCD fields $G$ alone. It is not appropriate to expand 
the effective action perturbatively since the
QCD fields, even in the vacuum condensate, are in general more 
significant than the masses of light and strange quarks.  When 
studying Eq.\,(\ref{PsiconEB}) one is  rapidly lead to the conclusion 
that the mass 
of {\it light} quarks is not the important scale in the problem 
at hand, since for $u,d,s$ quarks the glue condensate scale is greater 
than the quark mass. 

We note  that the substitution we have made in 
Eq.\,(\ref{twofields}) implies when effected in the highly
nonlinear function Eq.\,(\ref{PsiconEB}) that we have replaced 
\begin{equation}\label{factorG}
\langle G^{2n}\rangle \to (\langle G^{2}\rangle)^n
\end{equation}
which is correct under the tacit assumption that the fluctuations 
of the glue condensate can be understood as if generated by a 
background stochastic field~\cite{EdGM98,Dos87,Dos94}. This is than the 
picture of the true QCD vacuum that we consider to be very promising 
model of the complex reality and which is implied in the following.
The stochastic field fluctuations confine  the color charges of 
quarks and gluons, and determine  the confinement  size.

\subsubsection{Hadronic (Hyperfine) Structure}
Without going much into well studied  detail, the simple yet 
successful picture of hadronic structure is the 
`bag'-model. We view a hadron as comprising a spherical
confining volume of radius $R$ containing  $N$-quarks in
 a bound state of mass $M_N$. Neglecting here the quark mass compared to 
the zero-point $1/R$ energy the bound state mass is given by:
\begin{equation}\label{MN}
M_N=\frac{\kappa N}{R}+\frac{4}{3}\pi R^3 {\cal B} +H_{\rm I}\,.
\end{equation}
The residual QCD interaction term $H_{\rm I}$ comprises the quark-quark
interactions. The energy eigenvalue $\kappa/R$ of the relativistic 
quark wave function  with a boundary condition
enforcing a vanishing current through the surface gives
$\kappa=\kappa_c=2.04$.
The second term in Eq.\,(\ref{MN}) is the bag volume energy 
${\cal B}$, the latent (condensation) heat of the vacuum. The balance 
of forces requires that the stationary state is at
the minimum of $M_N(R)$ with respect to $R$, whence we
determine the mass of the  $N$-quark bound state:
\begin{equation}\label{MNN}
M_N= (\kappa N)^{3/4} 1.755 {\cal B}^{1/4}\qquad 
\mbox{for}\quad  H_{\rm I}=0\,.
\end{equation}

If we fit the nucleon mass $M_3=940$ MeV by Eq.\,(\ref{MNN}),
than we find ${\cal B}\to {\cal B}_{\rm N}^{1/4}=140$\,MeV, but 
working out the $\Delta$-mass, $M_3=1330$\,MeV one finds
${\cal B}_\Delta^{1/4}=190$\,MeV. 
These two values differ by factor 3.4 and thus 
we have to refine the approach. What is missing is 
the strong quark-quark interaction that is capable to
generate the hadronic multiplet 
mass splitting. The color-electric quark-quark 
interaction is comparatively weak as the overall color charge
of all hadronic states is zero, and only  particles comprising both 
light and heavy quarks have a sizable static electric color  charge 
density distribution. The color-magnetic hyperfine `spin-spin' 
interaction  term is~\cite{Cho74}:
\begin{eqnarray}
\label{hypef}
H_{\rm I}&=&\sum_{i>j\in h}\langle h | 
    \mu(r_{ij}) \frac{\lambda^i}{2}\cdot \frac{\lambda^j}{2}
    S^i\cdot S^j |h \rangle
\end{eqnarray}
Here $S=\sigma_a/2,\, a=1,2,3$ is the spin 
vector of the two interacting quarks $i,j$, 
written in terms of the Pauli spin matrices 
$\sigma_a$, and $\lambda_a/2, a=1\ldots8$ is 
the generator of the color non-Abelian SU(3) charge.
The dot product indicates summation over $a=1,2,3$ for SU(2) and 
$a=1\ldots8$ for SU(3). 

It turns out that all strongly 
interacting particles known (mesons
and baryons such as $\pi,\,\rho,\,N,\, \Delta,\,K,\,\Lambda$)
can  be consistently fitted using this color-magnetic interaction 
Eq.\,(\ref{hypef}), see~\cite{Deg75}. Let us recall here a 
ab-initio  fit to all strange and non-strange 
mesons and baryons,  allowing for massive strange quark variable 
scale~\cite{Aer84} which produces
${\cal B}_{\rm b}^{1/4}=170$ MeV. In this fit also the 
quark eigenenergy $\kappa$ was fitted, and 
the fitted value $\kappa=1.97$
is nearly the theoretical value $\kappa_{\rm th}=2.04$, the small 
difference between theory and fit expected to arise from the 
distortion of the wave-function due to the hyperfine 
color-magnetic interaction. The spectra of hadrons tell us that quarks 
live in a relativistic, confined orbital, which fact supports the 
QCD-quark-confinement-vacuum structure picture of the
 nucleon.

Another success of both the relativistic and the non-relativistic
quark bag model of hadrons is that we find that the most stable
 bound quark states have quantum 
numbers  seen in stable hadrons, and that the mass hierarchy of 
flavor  mass multiplets is correctly explained.  This  provides
firm evidence that the interaction between quarks is
obeying the rules of non-Abelian color-SU(3) algebra. To see this 
let us consider in more detail the situation for three quark states, 
i.e. baryons. We compute the
strength of the chromo-magnetic interaction $\mu_I^h$
in Eq.\,(\ref{hypef}), which we define as follows:
\begin{equation}
\label{mu1}
H_I=\delta\! E\,\mu_I^h\,;\qquad \mu_I^h\equiv
         -\sum_{i>j\in h_q}\langle h_q | 
         \frac{\lambda^i}{2}\cdot \frac{\lambda^j}{2}
        s^i\cdot s^j |h_q \rangle\,.
\end{equation}
Signs are chosen such that $\delta E\propto 1/R$ is a positive 
quantity. $\mu_I^h$ can be positive or negative, depending on
the quantum numbers of the bound state. 

To obtain the value of the matrix element $\mu_I^h$ in a given 
hadronic state we 
follow here the  permutation operator method~\cite{Joh75}. 
The exchange operator $P_{ij}$ of two quarks 
is composed of three factors for spin, color and flavor quantum 
numbers, assuming that in the ground state
the spatial wave function is identical for all quarks. 
For the totally antisymmetric quark state $|h_q\rangle$,
 for each quark pair $i,j$ we have thus 
\begin{equation}\label{PPP}
P_{ij}^cP_{ij}^s|h\rangle=-P_{ij}^f|h\rangle; 
\qquad P_{ij}^c=\frac{1}{3}+
        2\frac{\lambda^i}{2}\cdot \frac{\lambda^j}{2};
\qquad P_{ij}^s=\frac{1}{2}+2S^i\cdot  S^j
\end{equation}
The above explicit form of the permutation operators for the SU(2)-spin 
and SU(3)-color groups follows from the commutation properties of the 
generators $S=\sigma/2, \lambda/2$. Inserting Eq.\,(\ref{PPP}) into 
Eq.\,(\ref{mu1}) we obtain for the hyperfine interaction operator:
\begin{equation}\label{mu2}
\mu_I=\sum_{i>j}\left\{\frac{1}{4}P_{ij}^f +\frac{1}{24}
     +\frac14 \frac{\lambda^i}{2}\cdot \frac{\lambda^j}{2}
     +\frac{S^i\cdot  S^j}{6}\right\}\,,
     \end{equation}
and naturally $\mu_I^h=\langle \mu_I \rangle$\,.

To complete the evaluation of $\mu_I$ we need to commit 
ourselves to the symmetry group governing the flavor
exchange operator. Since strangeness is well distinguishable from
$u,d$ flavors one can take the point of view that we should use 
only the light quark SU(2) flavor group with:
\begin{equation}\label{Pf2}
P_{ij}^{f_2}=\frac12+2I^i\cdot I^j\,,
\end{equation}
and hence~\cite{Joh75}:
\begin{equation}\label{mu3}
\mu_I=\frac{n(n-1)}{12}+\sum_{i>j}\left\{
    \frac12 I^i\cdot I^j+\frac16 S^i\cdot S^j
+\frac14 \frac{\lambda^i}{2}\cdot \frac{\lambda^j}{2}\right\}
\,,
\end{equation}
where we have used $\sum_{i>j}=n(n-1)/2$. We than have to account
for the interaction of strange quarks with the non-strange quarks
separately. It turns out that the fitted~\cite{Aer84} chromo-magnetic 
interaction energy $\delta E_{qs}$  between strange and light quarks 
is half as large compared to $u,d$ interactions, and that 
strange-strange chromo-magnetic interaction  
$\delta E_{ss}$ is yet five times
weaker than the $u,d$ case. Such decrease of magnetic
interaction is consistent with the expectations that the more
massive quark has a smaller magnetic moment. While these 
results  support the need to separately treat 
the light and strange flavors, we 
note in passing that if we had taken $u,d,s$ quarks at par, 
than the larger SU(3)-flavor group would relate to flavor
symmetry. Accordingly in that case one should 
evaluate~\cite{BFP85} in Eq.\,(\ref{mu2}) 
the flavor exchange operator using the quadratic Casimir of the 
SU(3)-flavor group to represent the flavor permutation
operator in Eq.\,(\ref{PPP}).  We will not pursue this 
approach further here, which has merit when considering 
other aspects of hadronic spectra. 

Using the SU(2) flavor exchange operator we find that the 
chromo-magnetic inter\-action term of light quarks $u,d$
assumes the form:
\begin{equation}\label{mu3b}
\mu_I^{h_l}=\frac{n_l(n_l-6)}{12}+
    \frac14 I(I+1)+\frac1{12} S_l(S_l+1)+\frac14 C_2^l(p,q)
\,.
\end{equation}
where the index $l$ight reminds us that here we have only
accounted for  the light quark $u,d$ contribution and 
quantum numbers in all hadrons. In deriving Eq.\,(\ref{mu3b})
we have used some well known relations: 
$$2\sum_{i>j}I^i\cdot I^j=
  (\sum_iI^i)\cdot(\sum_jI^j) -\sum_i(I^i)^2\,, $$ $$
2\sum_{i>j}S^i\cdot  S^j=
     (\sum_i S^i)\cdot(\sum_j S^j) -\sum_i(S^i)^2\,$$
$$2\sum_{i>j}\lambda^i\cdot  \lambda^j=
    (\sum_i \lambda^i)\cdot (\sum_j \lambda^j)-
    \sum_i\lambda^i\cdot \lambda^i $$
and
$$
(\sum_i I^i)^2|h\rangle =I(I+1)|h\rangle\,,\qquad
(\sum_i s^i)^2|h\rangle =S(S+1)|h\rangle\,.
$$
given the total spin and isospin operators introduced 
implicitly above. Everything works 
in the same way for the color group SU(3), 
except that the individual quadratic Casimir 
operator has a different eigenvalue to keep in mind. 
In the fundamental representation
$$
C_2(\underline{3})\equiv\sum_{a=1}^8
     \left(\frac{\lambda_a^i}{2}\right)^2=\frac43\,,
$$
which is just the inverse of 3/4, the eigenvalue of the quadratic
Casimir operator for the SU(2) in the fundamental 
representation \underline{2}. For the 
composite state $|h\rangle $ of particles $q$ and antiparticles $\bar q$
in fundamental representation we have:
$$
\sum_{a=1}^8\left(\sum_i \frac{\lambda_a^i}{2}\right)^2|h\rangle =
C_2(p,q)|h\rangle\,,
$$
where the values of $(p,q)$ for a few representations are:
\underline{1}=(0,0), \underline{$\bar 3$}=(0,1), 
\underline{$ 3$}=(1,0), 
\underline{$ 6$}=(2,0), \underline{$\bar 6$}=(0,2),
\underline{8}=(1,1), \underline{10}=(3,0), etc.
More generally, the dimension of the representation and the 
Young-tableaux numbers $p,q$ are related by:
$$d(p,q)=\frac12 (p+1)(q+1)(p+q+2)\,.$$
The quadratic Casimir operator in terms of $p,q$ assumes the values
$$
C_2(p,q)=p+q+\frac{p^2}{3}+\frac{pq}{3}+\frac{q^2}{3}\,.$$
Explicitly:
$$
C_2(\underline{1})=0\,,\ \
C_2(\underline{3})=C_2(\underline{\bar3})=\frac43\,,\ \
C_2(\underline{6})=C_2(\underline{\bar6})=3\frac13\,,\ \
C_2(\underline{8})=3\,.
$$

It is with great satisfactions that we note that
 Eq.\,(\ref{mu3b}) has a minimum 
for $n_l=3$, corresponding to the baryon system, and that overall
the chromo-magnetic interaction energy factor is negative 
only for the nucleon state among the three quark states. 
The splitting of $N$ and $\Delta$ (difference between 
$I=3/2, S=3/2$ and $I=1/2, S=1/2$ states) $M_\Delta-M_N=\delta E$
thus leads to\footnote{
When fitting the mass of the $\Delta$ one has to 
note that the coupling to a strong decay channel also shifts the 
quark based hadron mass downward yielding the physical mass, 
and hence the quark-based portion of the $\Delta$
mass is to be taken at $M_\Delta+\Gamma_\Delta$
} $\delta E\simeq -400$MeV, which is consistent with the 
splitting $M_\Sigma-M_\Lambda=(3/16)\, \delta E$ 
(difference between $I=0$ 
and $I=1$ states).  

These are very persuasive examples of the
successes of the bag model of quark structure 
of hadrons, and prove  that 
the hadron masses are properly 
described by the color magnetic interaction. 

\subsubsection{Limitations of the quark-bag picture}
But if Eq.\,(\ref{hypef}) works very well for the fundamental multiplets 
of baryons and mesons, why is it that it also predicts unwanted
states such as baryonium $qq$--$\bar q \bar q$, the $qqs$--$qqs$ di-baryons
and many more, not seen? Some of these exotica are quite stable: which
multi-quark states are most bound can be deduced from the 
general bonding (Hund) rule of quark physics~\cite{Jaf77}:
\begin{itemize}
\item the  quarks and antiquarks are separately in the largest possible
representation of color and spin, and
\item the total state is in the smallest possible representation of 
spin an color. 
\end{itemize}
Thus the  state with the structure 
$[(uds)_{8_c,4_s}\otimes(uds)_{8_c,4_s}]_{0_c,0_s}$
where the sub-indices refer to color and spin multiplicities of the 
representations is the most bound exotic 6 quark state. 
Such a hadron has not been found despite literally twenty different 
experimental searches. So probably not all is well with our current 
understanding of quark structure in hadrons and strong interactions. 
It could be that  the small size of normal
hadrons may not yet require a fully dissolved vacuum structure, 
while the larger  exotica do require full structure  dissolution,
In such a case there is additional vacuum energy cost required 
to form these larger exotic states, a fact which would induce considerable 
instability by hadronic dissociation. The hope is that the study of 
really large chunks of quark matter we call quark-gluon
plasma will provide a shortcut to a resolution of  this problem.

Let us hence make a step back and look 
again at the hadron structure fit~\cite{Aer84}
--  if this fit is so successful, what does it tell us about the 
volume energy, which is the `condensation' latent energy of the
 vacuum? The fitted 
value ${\cal B}_{\rm b}^{1/4}=170$ MeV, which we also can 
express as ${\cal B}_{\rm b}=0.1 $Gev/fm$^3$ is  relatively
small. The magnitude derives  from the fact that the
quark-related energy component in the hadron 
is 3/4 of the total, as can be seen solely on 
dimensional grounds inspecting Eqs.\,(\ref{MN},\ref{MNN}), 
since only terms changing with $R$ as 
$R^3$ and $1/R$ appear. The volume energy component in
the nucleon is thus about $m/4=235\equiv {\cal B}_{\rm b}V_h$ MeV. 
Only if we could make hadron volume $V_h$ very small, 
 the volume energy could come out big. However
the elastic electromagnetic form factors and other electromagnetic
properties of hadrons  indicate that the Maxwell-charge
distribution inside hadron is not less than 0.7--0.8 fm
in size. The above considered value of the  volume energy leads to
a quark confinement volume with radius $R=0.83$\,fm, for protons,
compared with the charge radius 
 $<r^2>=(0.86\,\mbox{fm})^2 $. Clearly, there is not much  room 
for change and surely we will not be able to find 
${\cal B}={\cal O}(1)$ GeV/fm$^3$, a value motivated by 
numerical simulations of lattice QCD~\cite{lattice}.

\section{Quark-Gluon Plasma} \label{QGPsec}
\subsubsection{Relativistic heavy ion experiments}
In order to form  relatively large quark-gluon filled space-time regions 
the best tool we have today are large nuclei. 
These nuclei are made to collide at very high energies, many times
higher than the rest mass of particles involved. The relativistic 
energy is required to produce regions of space filled with movable
color charges of quarks and gluons, the quark-gluon plasma (QGP). 
Primarily because the identification of this new, locally deconfined 
form of matter is difficult, the object we search for
seems to be the never-to-be-found nuclear Holy Grail. However, 
this deconfined QGP state must in principle exist if 
QCD is the true theory of strong
interactions: such a free quark-gluon phase 
is clearly seen in numerical studies of 
QCD within the lattice gauge theory (LGT) simulations,
though the details of the quark structure of the deconfined
state remain today somewhat obscured by the difficulties that
one encounters in numerical treatment of Fermions~\cite{lattice}.

The study of highly excited and dense hadronic matter by means of
ultra-relativistic nuclear collisions is a relatively novel,
interdisciplinary area of
research in rapid experimental and theoretical evolution. It is
closely related to the fields of nuclear and
particle physics and, accordingly, our material encompasses aspects
of both these wide research areas. Among `consumers' of the
material presented here we also encounter researchers in the fields
of astrophysics and cosmology.
The idea to test this is to squeeze and compress large nuclei, 
such that individual  nucleons are made to fuse into a new common 
structure comprising freely moving quarks.

The initial experimental programs were launched nearly 30 years ago 
at the Lawrence Berkeley Laboratory (LBL) at Berkeley,
USA, and at the Dubna Laboratory (Russia) for lighter ions. 
Another 1-2 GeV facility,
SIS, of comparable physics perspective  has come on line in recent years
 at GSI in Darmstadt, Germany. A greater beam 
intensity and new detector
technologies at SIS allow to explore systematically phenomena that
could not be studied at the BEVALAC.
Studies of properties of excited nuclear matter 
performed at these facilities
display particle spectra `temperatures' of the order
of 100 MeV, accompanied by relatively
abundant pion production. 

These first advances where made with
relatively modest energies, and the results had mainly bearing on
the properties of the nuclear matter close to the conditions
encountered in supernova explosions. However, they demonstrated
the possibility to study the properties of compressed and excited 
nuclear matter in laboratory. 
Together with the theoretical work, these results implied that QGP
could be in reach of existing accelerator facilities. 
Thus a search for the point of
transition from the hadronic gas phase of locally confined nucleons, 
mesons,  etc,  to the deconfined QGP phase that gave birth to the
development of the present day research program, in USA at BNL
(Brookhaven National Laboratory, Upton -- Long Island, New York), 
and in Europe at CERN (European Nuclear Physics Research Laboratory,
Geneva).  The first Oxygen beam at 60~A~GeV was
extracted from the SPS (Super Proton Synchrotron) 
accelerator at CERN and met target in the late
fall of 1986, about the same time as BNL started its experimental
program at AGS (Alternate Gradient Synchrotron) accelerator
with 15~A~GeV Silicium ions. 
Very soon after, the energy of
the SPS beam could be increased to 200~A~GeV and a Sulphur ion
source was added. 
In order to allow experiments to study the relatively large volumes
and longer lifetimes expected in dense matter formed in collision
of heaviest nuclei, an upgrade of SPS injection system was
carried out, which allowed to accelerate the Lead ions to 158~A~GeV in
the Fall of 1994. At BNL, a similar development allowed to accelerate 
Gold beams  to 11~A~GeV at the AGS. 

Today we are repositioning our interests around 
two new experimental facilities: RHIC and LHC.
The RHIC (Relativistic Heavy Ion Collider)  at BNL, is slated for 
physics progress in 1999,  with colliding nuclear 
beams at up to 100 A GeV, thus
allowing exploration of a entirely new domain of energy, 10 times
greater than currently available at SPS. The LHC (Large Hadron
Collider) which is being developed at CERN, should be commissioned
at midpoint of the first decade of XXI century, and it will devote an
important part of its beam time to the acceleration of nuclear
beams. The final energy available at the 
LHC is expected  to be a factor 30 or more higher
compared to RHIC.

Many experiments are in progress both at CERN and BNL and 
our present interest addresses primarily results obtained
at the higher CERN energies, involving strange particle
production. Several experiments  comprise similar `crews' and we
will group their names together. This leaves
us with two major lines of approach:
 NA35/NA49 and WA85/WA94/WA97/NA57. While the NA35 etc groups are
interested in obtaining full phase space coverage and use 
detectors such as streamer chambers and time projection chambers, 
the WA85 etc groups concentrate on a small window of phase space
`opportunity' with a spectrometer, which allows to identify
and observe rarely produced multi-strange particles.  
Another series of experiments which is of some importance 
for our study of heavy flavor quarks is NA38/NA50/NA51 series,
which observes by means of their dimuon decays the production of
vector mesons. 

The first conclusion one can draw from the recent 
results including the  collisions of Pb--PB nuclei involving projectiles
colliding with a laboratory fixed target at 158 A GeV is that 
there is  a localized space-time region in which high concentration
of matter and energy is reached: we refer to this object as 
`fireball'. 
We believe that local thermal equilibrium is established 
within the volume occupied by the fireball, and thus particles
emerge with spectra characteristic of the surface temperature
and velocity.

\subsubsection{Probes of dense hadronic matter}
Since in the collision of large  nuclei,  
the highly dense state is 
formed for a rather short time of magnitude $2R/c$, 
where $R$ is the nuclear
radius, one of the major challenges 
has been to identify suitable physical 
observable of deconfinement. 
This difficult problem of detecting 
reliably the formation of an 
unknown  phase of matter, existing only as short 
as $0.5\cdot 10^{-22}$sec, has not been
completely resolved today. 
The electromagnetic probes involving directly
produced photons and dileptons are witnesses 
to the earliest moments 
of the reaction, but their production rates are in general very small,
the  experimental $\gamma$
yield is dominated by the secondary processes $\pi^0\to \gamma+\gamma$.
The dilepton yield is also mostly resulting from meson decays, but 
there are small kinematic regions where this background yield is very small. 
Moreover, the dilepton spectrum gives interesting insights about the 
vector meson yields and their variation with experimental conditions. We
will briefly address here  this very interesting observable.

The principal observable of the relativistic nuclear 
collision process are spectra of hadronic (strongly interacting)
particles. Their multiplicity is growing rapidly with both 
energy and size of the projectile-target nuclei. 
Aside of the single particle spectra
one can also relatively easily measure two 
particle correlations. The  HBT correlation measurement
is widely used to determine the geometric properties of the central
fireball, and the results agree with a reaction picture
between the nuclei relying on geometric considerations.
However, this approach is still plagued by 
fundamental and unresolved issues. 
Work is progressing rapidly in this field and which is 
addressed in full in this volume by Baym and Heinz~\cite{Bay97}.

Strangeness~\cite{Raf82} (and at RHIC/LHC also charm) 
and entropy~\cite{entropy} are good hadronic 
observable since both will
 be preserved  by `reasonable' evolution scenarios
of the dense matter fireball: the melted QGP state is in general 
entropy richer than the frozen HG phase. 
Once entropy has been generated, it cannot be lost, an entropy 
excess accompanies QGP formation. When abundant particle production 
is possible, this entropy excess is seen as an enhancement in the total 
hadronic particle multiplicity with  each 
(relativistic) meson carrying about 4 units of entropy
out of the interaction region.
Similarly, strangeness is in general more abundant 
in QGP than HG phase~\cite{Raf82}, and 
it is not re-annihilated in rapid decomposition of the 
dense matter state~\cite{KMR86}. 
It has become a key  diagnostic tool of dense hadronic 
matter because:\\
\indent {1)} particles containing  strangeness are produced more 
abundantly in relativistic nuclear collisions than it could be 
expected based on simple scaling of $p$--$p$ reactions;\\ 
\indent {2)} all strange quarks  have to be
made, while light $u$, $d$  quarks are  also
brought into the reaction by the colliding nuclei;\\ 
\indent {3)}
because there are many different strange particles, we have a
very rich field of observable with which it is possible to explore
diverse properties of the source;\\ 
\indent {4)} theoretical calculations suggest that glue--glue
collisions in the QGP provide a sufficiently fast and
thus by far, a unique mechanism leading to an explanation of
strangeness enhancement. 

There are two generic flavor 
observable (strangeness and charm) which we  study  
analyzing experimental data, and we introduce these here,
without an effort to `orthogonalized', {\it i.e.}, make 
them independent of each other:
\begin{itemize}
\item {absolute yield} of strangeness/charm\\ 
Once produced in hot and dense hadronic matter, {\it e.g.}, 
the QGP phase, strangeness/charm is
not re-annihilated in the evolution of the deconfined state towards
freeze-out, because in the expansion and/or cooling process the rate of 
production/annihilation rapidly diminishes and becomes negligible. 
Therefore the flavor yield is characteristic of the
initial, most extreme conditions. 
 \item {phase space occupancy  $\gamma_{i}$}\\
$\gamma_{i}$ describes how close the flavor  yield per 
unit of volume ($i={\rm u,d,s}$, and in some cases charm 
$i={\rm c}$ occupancy
will be also considered) comes to the chemical equilibrium expected; 
$\gamma_i$  impacts strongly the distribution of 
flavor among final state  hadronic particles.
\end{itemize}
Because of the high density of the QGP phase, the 
phase space occupancy $\gamma_{i}$ 
can saturate rapidly, and thus particle abundances 
will emerge from a chemically equilibrated $u$, $d$, $s$ phase, which is
hardly imaginable for conventional reaction mechanisms.
Because entropy and
strangeness are enhanced in a similar way in QGP, the specific yield of
strangeness per particle produced is not a good quantity to
use when searching for the deconfined state. Many other strategies
are available, of which we favor measurement of the specific 
entropy yield per participating baryon, accompanied 
by a study of relative 
strange antibaryon yields, involving particle ratios such 
as $\overline{\Lambda}/\bar p$~\cite{Raf82}. It is remarkable that
the pertinent results obtained for S--Pb collisions 
by the NA35 collaboration~\cite{NA35pb} have 
shown the QGP related enhancement. 

\subsubsection{Recent results on strangeness}
The possibility that strange particle anomalies seen in recent
years at SPS  in Sulphur  induced reactions on heavy
nuclei are  arising in consequence to the formation of a 
deconfined QGP phase
has stimulated the intense continuation of the experimental 
research program in the considerably more difficult, 
high particle  multiplicity environment arising in Pb induced 
reactions, which are presently possible at 158A GeV.

\begin{figure}[!htb]
\vspace*{5cm}
\centerline{\hspace*{0.6cm}
\psfig{width=14cm,figure=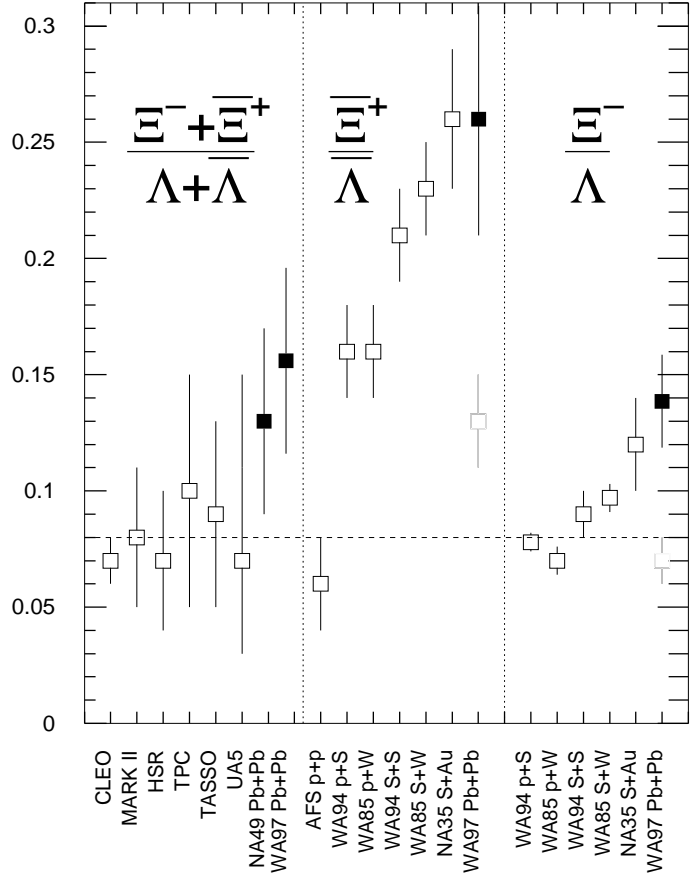}
}
\vspace*{-2.5cm}
\caption{
Sample of World results (as `function' of experiment name
for yields involving ratio of
strange to non-strange quarks in baryons.
 Dark squares: recent Pb--Pb results.
\label{RSS}
}
\end{figure}

The experiment WA97~\cite{WA97}  has further reported several specific 
strange baryon and  anti\-baryon  ratios from 
Pb--Pb collisions at 158A GeV, comprising 30\% of inelastic 
interactions. All ratios are obtained in an overlapping kinematic 
window corresponding effectively to
transverse momentum $ p_\bot>0.7$ GeV, within the central 
rapidity region $ y\in y_{cm}\pm0.5$. They have been corrected for weak
interactions cascading  decays. The experimental values are:
\begin{eqnarray}\label{rat1}
R_\Lambda\!=\!\frac{\overline{\Lambda}}\Lambda\! =\! 0.14\pm 0.03, \ \
R_\Xi\!=\!\frac{\overline{\Xi}}\Xi        \! =\! 0.27\pm 0.0,\ \
R_\Omega\!=\!\frac{\overline{\Omega}}\Omega  \! =\! 0.42\pm 0.12,\ \ 
\end{eqnarray}
\vspace*{-0.6cm}
\begin{eqnarray}
R_{\rm s}^{\rm p}=\frac\Xi\Lambda=0.14\pm 0.02\,,  &&\quad   
R_{\bar {\rm s}}^{\rm p} =\frac{\overline\Xi}{\overline\Lambda} = 0.26\pm 0.05\,,\\
\ \nonumber \\
{R'_{\rm s}}^{\rm p}=\frac\Omega\Xi =0.19\pm 0.04\,,  &&\quad
{R'_{\bar {\rm s}}}^{\rm p}= \frac{\overline\Omega}{\overline\Xi}  = 0.30\pm 0.09\,.
\end{eqnarray}
Here, the lower index s,
resp. $\bar {\rm s}$, reminds us that the ratio measures the density of
strange, resp. anti-strange, quarks relatively to light quarks.
The upper index $p$ indicates that the ratio is taken
within a common interval of transverse momenta (and not common
transverse mass). We compare $R_{\rm s}^{\rm p},\, R_{\bar {\rm s}}^{\rm p}$   
results with 
earlier measurements in Fig.\,\ref{RSS}. The strange antibaryon enhancement
effect is re-confirmed in the Pb--Pb data, and we see that there is no major 
change of this result, which determines the phase space occupancy 
of strangeness, as we move from S--S  or S--W/Pb results to Pb--Pb 
results. 
There is agreement between WA97 and NA49 
on the value of $R_\Lambda$,
even though the data sample of NA49 is taken for more central trigger,
constrained to as few as 4\% of most central  collisions. 
The cuts in $p_\bot$ and $ y$
are nearly identical in both experiments. From Fig.\,3 in 
~\cite{NA49B}, we obtain the value 
$R_\Lambda=0.17\pm0.03$, which we shall combine with the value given by
WA97 and we thus take in out data fit:
\begin{eqnarray}\label{rat2}
R_\Lambda=\frac{\overline{\Lambda}}\Lambda = 0.155\pm 0.04\,.
\end{eqnarray}

\section{Strangeness Production}\label{sprod}
\subsubsection{The coupling constant of strong interactions}
The observed strange particles are presumably born in a 
deconfined phase which can make strange quark pairs effectively.
It turns out that the production of strangeness is 
originating primarily in gluon-gluon collisions.
We use  two particle collision 
processes  to evaluate thermal flavor 
production in QGP, as described in~\cite{Rio95}. However, since
this report we have realized~\cite{impact96,acta97} that 
recent precise measurements 
of the strong interaction coupling constant $\alpha_{\rm s}$ 
allow us to eliminate a lot of arbitrariness from the earlier 
calculations regarding the strength of the collision cross sections.
The experimental measurements of $\alpha_{\rm s}$ occurred at 
the scale of the the Z$_0$ mass, thus 92 GeV, and hence one has to 
use the powers of the renormalization group of QCD to evolve these
results towards the here relevant domain. In the process we 
also allow for the evolution of the strange quark mass, which is
also significant and is driven by the evolution of the 
coupling constant. We will now briefly explain how
this is accomplished. 

To determine the two QCD parameters required, we will use the
renormalization group functions $\beta$ and $\gamma_{\rm m}$:
\begin{equation}\label{dmuda}  
\mu \frac{\partial\alpha_{\rm s}}{\partial\mu}
=\beta(\alpha_{\rm s}(\mu))\,,\qquad
\mu {\frac{\partial m}{\partial\mu}} =-m\,
\gamma_{\rm m}(\alpha_{\rm s}(\mu))\,.
\end{equation}
For our present study we will use the perturbative power expansion 
in $\alpha_{\rm s} $:
\begin{equation}\label{betaf}
\beta^{\rm pert}-\alpha_{\rm s}^2\left[\ b_0
   +b_1\alpha_{\rm s} +\ldots\ \right] \,,\quad
\label{gamrun}
\gamma_{\rm m}^{\rm pert}=\alpha_{\rm s}\left[\ c_0
+c_1\alpha_{\rm s} + \ldots\ \right]\,,
\end{equation}
For the SU(3)-gauge theory with $n_{\rm f}$ Fermions the first two
terms (two `loop' order) are renormalization scheme independent,
the three loop level $b_2$ is known in both MS (minimal subtraction) 
and $\overline{\mbox{MS}}$ (modified minimal subtraction)
renormalization schemes~\cite{QCD95,SS96}, while $c_2$, so far we 
could ascertain, was only derived in the MS scheme. In any case 
we will restrict our study to the two loop order which does
not require renormalization scheme improvements of the form of
the two body cross sections.  To this order we have:
\begin{eqnarray}
b_0=&\!\!\!  \displaystyle\frac{1}{ 2\pi}
  \left(11-{2\over 3}n_{\rm f}\right)\,,\quad
b_1=&\!\!\! \frac{1}{4\pi^2}\left(51-{19\over 3}
        n_{\rm f}\right)\,,\\ \ \nonumber\\
 c_0=&\!\!\! \displaystyle\frac{2}{\pi}\,,\hspace*{2.9cm}
c_1=&\!\!\!\frac{1}{12\pi^2}
        \left(101-{10\over 3}n_{\rm f}\right)\,.
\end{eqnarray}

The number  $n_{\rm f}$ of Fermions that can be excited, depends
on the energy scale $\mu$. We have implemented this using the
exact phase space form appropriate for the terms linear
in $n_{\rm f}$
\begin{eqnarray}\label{nfs}
n_{\rm f}(\mu)=2+\sum_{i=s,c,b,t}\sqrt{1-\frac{4m_i^2}{\mu^2}}
  \left(1+\frac{2m_i^2}{\mu}\right)\Theta(\mu-2m_i)\,,
\end{eqnarray}
with $m_{\rm s}=0.16\,{\rm GeV},\,m_{\rm c}=1.5\,{\rm GeV},\,
m_b=4.8$\,{\rm GeV}. We checked that there is very minimal impact
of the running of the masses in Eq.\,(\ref{nfs}) on the final
result, and will therefore not introduce that `feed-back' effect
into our current discussion. The largest influence on our solutions
comes here from the bottom mass, since any error made at about 5
GeV is
amplified most. However, we find that this results in a scarcely
visible change even when the mass is changed by 10\% and thus 
one can conclude that the exact values of the masses and the 
nature of flavor threshold is at present of minor importance in our
study. 
\begin{figure}[!htb]
\vspace*{-1.6cm}
\centerline{\hspace*{0.6cm}
\psfig{width=12cm,figure=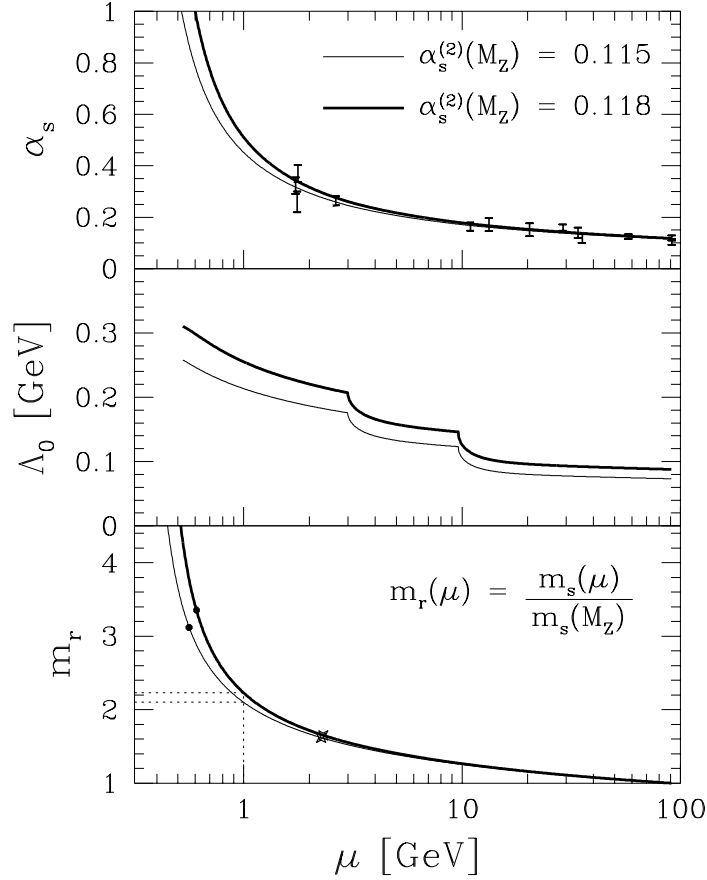}}
\vspace*{0.2cm}
\caption{\small
\rm$\alpha_{\rm s}(\mu)$ (top section); the equivalent 
parameter $\Lambda_0$ (middle section) and
$m_{\rm r}(\mu)=m(\mu)/m(M_{Z})$
(bottom section) as function of energy scale $\mu$. 
Initial value $\alpha_{\rm s}(M_{Z})=0.118$ (thick solid lines)
and $\alpha_{\rm s}(M_{Z})=0.115$ (thin solid lines).
In lower section the dots indicate the strangeness pair production
thresholds for $m_{\rm s}(M_{Z})=$~90~MeV, while crosses indicate
charm pair production thresholds for $m_{\rm c}(M_{Z})=$~700~MeV.
\label{fig-a1}}
\end{figure}
We show the result of numerical integration for $\alpha_{\rm s}$ in
the top portion of Fig.\,\ref{fig-a1}. First equation in
(\ref{dmuda}) is numerically integrated  beginning with an initial
value of $\alpha_{\rm s}(M_Z)$. We use in this report
the August 1996 World average\,~\cite{Sch96}: 
$\alpha_{\rm s}(M_{{Z}})=0.118$ for which the estimated error is 
$\pm 0.003$\,. This value is sufficiently precise to 
reduce this uncertainty below that has limited our earlier
study\,~\cite{impact96}. In addition, the thin solid lines
present results for $\alpha_{\rm s}(M_{{Z}})=0.115$\, till recently
the preferred result in some analysis, especially those at lower
energy scale. As seen in Fig.\,\ref{fig-a1}, the variation of
$\alpha_{\rm s}$ with the energy scale is substantial, and in
particular we note the rapid change at and below $\mu=1$ GeV, where
the strange quark flavor formation occurs in hot QGP phase formed
in present day experiments at 160--200 A GeV (SPS-CERN). Clearly,
use of constant value of $\alpha_{\rm s}$ is hardly justified, and
the first order approximation often used:
\begin{equation}\label{Lambdarun}
\alpha_{\rm s}(\mu)\equiv
  \frac{2b_0^{-1}(n_{\rm f})}{\ln(\mu/\Lambda_0(\mu))^2}\,,
\end{equation}
can be seen to be not a good approximation 
till rather high $\mu=50$ GeV scale is reached:
we insert the numerically computed value of 
$\alpha_s$ into Eq.\,(\ref{Lambdarun}) and obtain $\Lambda_0(\mu)$ 
which would yield the given $\alpha_s(\mu)$. This procedure 
leads to a strongly scale dependent $\Lambda_0(\mu)$ shown in the
middle  section of  Fig.\,\ref{fig-a1}. 

Given $\alpha_{\rm s}(\mu)$, we can
integrate the running of the quark masses, the second  equation in
(\ref{dmuda}).  Because the running mass equation 
is linear in $m$, it is possible to determine the universal 
quark mass scale factor
\begin{equation}
m_{\rm r}=m(\mu)/m(\mu_0)\,.
\end{equation}
Since  $\alpha_{\rm s}$ refers to  the scale of $\mu_0=
M_Z$, it is a convenient reference point also for quark masses.  
As seen in the bottom portion of Fig.\,\ref{fig-a1},
the change in the quark mass factor is highly relevant in the study
of strangeness and  charm production,
since it is driven by the rapidly changing
$\alpha_{\rm s}$ near to $\mu\simeq 1$~GeV.
For each of the different functional dependencies
$\alpha_{\rm s}(\mu)$ we obtain a different function
$m_{\rm r}$. The significance of the running of the charmed
quark mass cannot be stressed enough, especially for thermal charm
production occurring in foreseeable future experiments well below
threshold, which makes the use the exact value of 
$m_{\rm c}$ necessary.
 
Given these results, we find that for $\alpha_{\rm s}=0.118$ and
$m_{\rm s}(M_{{Z}})=90\pm18$~MeV a low energy strange quark mass
$m_{\rm s}(1\mbox{\,GeV})\simeq 200\pm 40$ MeV, in the middle of
the standard range $100<m_s$(1\,GeV) $<$ 300 MeV. Similarly we
consider $m_{\rm c}(M_{{Z}})=700\pm50$~MeV, for which value 
we find  the low energy mass $m_c(1\mbox{\,GeV})\simeq 1550\pm110$
MeV, at the upper (conservative for particle production yield) end
of the standard range $1<m_c$(1\,GeV) $<1.6$ GeV. The energy 
threshold for pair production $E^{\rm th}_i,\,i=$ s, c, 
has to be determined in a separate calculation since it is
related to the scale of energy required 
for the production of two zero momentum particles
$E_{\rm th}=2m(\mu=E_{\rm th})$, and thus:
\begin{equation}\label{dispersion} 
E_i^{\rm th}=2m_i(M_{{Z}})m_{\rm r}(E_i^{\rm th})\,.
\end{equation}
This effect stabilizes strangeness production cross section in the
infrared: below $\sqrt{\rm s}=1$ GeV the strange quark mass
increases rapidly and the threshold mass 
is considerably greater than  $m_{\rm s}$(1 GeV).
We obtain the threshold values $E_{\rm s}^{\rm th}=605$ MeV
for  $\alpha_{\rm s}(M_Z)=0.118$ 
and $E_{\rm s}^{\rm th}=560$ MeV for $\alpha_{\rm s}(M_Z)=0.115$. 
Both values are indicated by the black dots in 
Fig.\,\ref{fig-a1}.  For charm, the running mass effect has 
the opposite effect:  since the mass of charmed 
quarks is listed in tables
for $\mu=1$ GeV, but the here explored 
value of the mass is above 1 GeV, the charm
production threshold turns out to be smaller than expected.
There is little uncertainty with regarding the coupling constant;
for $m_{\rm c}(M_Z)= 700$ MeV the
production threshold is found at $E_c^{\rm th}\simeq 2.3$ GeV
rather than $2\cdot 1.55= 3.1$ GeV, the value we would 
have obtained using $m_c(1\mbox{\,GeV})$ in this case.
This reduction in threshold energy enhances strongly 
the thermal production of charm at `low' temperatures $T\simeq
250$ MeV, where it is suppressed exponentially by a factor 
$\propto\exp(-E^{\rm th}/T)$.
 
\subsubsection{QCD strangeness production processes}

The generic angle averaged two particle cross sections for (heavy)
flavor production processes
$g+g\to f+\bar f $ and $ q+\bar q\to f+\bar f\,,$ are among
classic results of QCD and are given by
\begin{eqnarray}
\bar\sigma_{gg\to f\bar f}(s) &=&
   {2\pi\alpha_{\rm s}^2\over 3s} \left[
\left( 1 + {4m_{\rm f}^2\over s} + {m_{\rm f}^4\over s^2} \right)
{\rm tanh}^{\!-1}W(s)-\right.\nonumber\\
&&\hspace*{3cm}\left.-\left({7\over 8} + {31m_{\rm f}^2\over
8s}\right) W(s) \right]\,,\label{gl}\\
\bar\sigma_{q\bar q\to f\bar f}(s) &=&
   {8\pi\alpha_{\rm s}^2\over 27s}
   \left(1+ {2m_{\rm f}^2\over s} \right) W(s)\,,
\label{gk}
\end{eqnarray}
where $W(s) = \sqrt{1 - 4m_{\rm f}^2/s}$\,, and both the QCD
coupling constant $\alpha_{\rm s}$  and flavor quark mass $m_{\rm
f}$ will be in this work the running QCD parameters. 
In this way a large number of even-$\alpha_{\rm s}$
diagrams contributing to flavor production  is accounted for.
 
Other processes
in which at least one additional gluon is present are not 
within the present calculational scheme. While only in
very high density environment we could imagine relevant
contributions from three body initial state collisions, emission
of one or even several soft gluons in the final state could be
relevant,
thus this subject area will be surely revisited in the future. 
We note that a process in which a massive `gluon', that is a
quasi-particle with quantum numbers of a gluon, decays into a
strange quark pair, is partially included in our work.
At the present time we do not
see a systematic way to incorporate any residue of this and other
effects, originating in matter surrounding the microscopic
processes, as work leading to understanding of renormalization
group equations in matter (that is renormalization group 
at finite temperature and/or
chemical potential) is still in progress\,~\cite{Elm95}. 
 
The master equation for flavor production,
allowing for the detailed balance reactions, thus
re-annihilation of flavor, is:
\begin{equation}\label{dNsdt}
{{dN_{i}(t)}\over {dt}} = 
V(t)A_{i}\left[1-\gamma_{i}^2(t)\right]\,.
\end{equation}
where $A_i$ is the invariant rate per unit volume and time,
that is the thermal average of the production cross section:
\begin{eqnarray}
A_{i}&\!\!\!\equiv&\!\!\! A_{gg}+A_{u\bar u}+A_{d\bar
d}+\ldots\nonumber\\
&\!\!\!=&\!\!\!\int_{4m_{i}^2}^{\infty}ds
2s\delta (s-(p_1+p_2)^2)
\int{d^3p_1\over2(2\pi)^3E_1}\int{d^3p_2\over2(2\pi)^3 E_2}
\nonumber\\ 
&&\hspace*{0.8cm}\times
\left[{1\over 2} g_g^2f_g(p_1)f_g(p_2)
\overline{\sigma}_{gg}(s) + n_{\rm f}g_q^2 f_q(p_1) 
f_{\bar q}(p_2)\overline{\sigma}_{q\bar
q}(s)+\ldots\right]\,.\label{qgpA} 
\end{eqnarray}
The dots indicate that other mechanisms may contribute to
flavor production. The particle distributions $f$ are in our
case thermal Bose/Fermi functions, and $g_{\rm q}=6,\,g_{\rm g}=16$\,.
 For strangeness
production $n_{\rm f}=2$, and for charm production $n_{\rm f}=3$\,,
will be used.

We introduce also the flavor production relaxation time constant:
\begin{equation}\label{tauss}
\tau_{i}\equiv {1\over 2}{\rho_{i}^\infty(\tilde m_{i})
                 \over{(A_{gg}+A_{qq}+\ldots)}}\,. 
\end{equation}
Here $\tilde m_{i}$ is the mass at the scale of energy under study;
we recall that the equilibrium distribution is result of Boltzmann 
equation description of two body collisions. Thus the mass arising
in the equilibrium density $\rho_{i}^\infty$ in
Eq.\,(\ref{tauss}) is to be taken at the energy scale of the
average two parton collision. We adopt for strangeness a fixed
value ${\tilde m}_{\rm s}=200$~MeV, and  also ${\tilde m}_{\rm c}=
1,500$~MeV, and observe that in the range of temperatures here 
considered the precise value of the strange quark mass is
insignificant, since the quark density is primarily governed by the
$T^3$ term in this limit, with finite mass correction being ${\cal
O}$(10\%). The situation is less clear for charm relaxation, since
the running of the mass should have a significant impact.

\begin{figure}[!htb]
\vspace*{-0.8cm}
\centerline{\hspace*{-0.5cm}
\psfig{width=11.5cm,figure=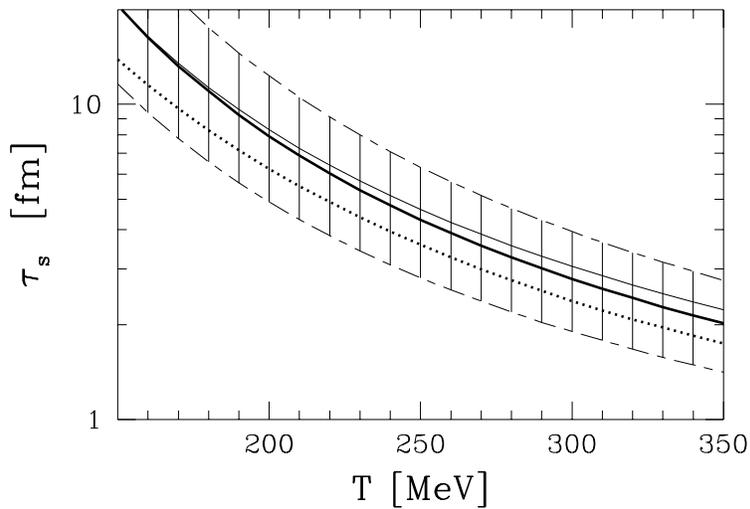}}
\vspace*{-0.3cm}
\caption{\small \label{figTaussrun}
QGP strangeness relaxation time, for $\alpha_{\rm s}(M_{Z})=0.118$,
(thick line) and = 0.115 (thin line); $m_{\rm s}(M_{{Z}})=90$~MeV.
Hatched areas: effect of variation of strange quark mass by 20\%.
 Dotted: comparison results for fixed  
$\alpha_{\rm s}=0.5$ and $m_{\rm s}=200$ MeV.
}\end{figure}
\begin{figure}[!htb]
\vspace*{-0.3cm}
\centerline{\hspace*{-0.5cm}
\psfig{width=11.5cm,figure=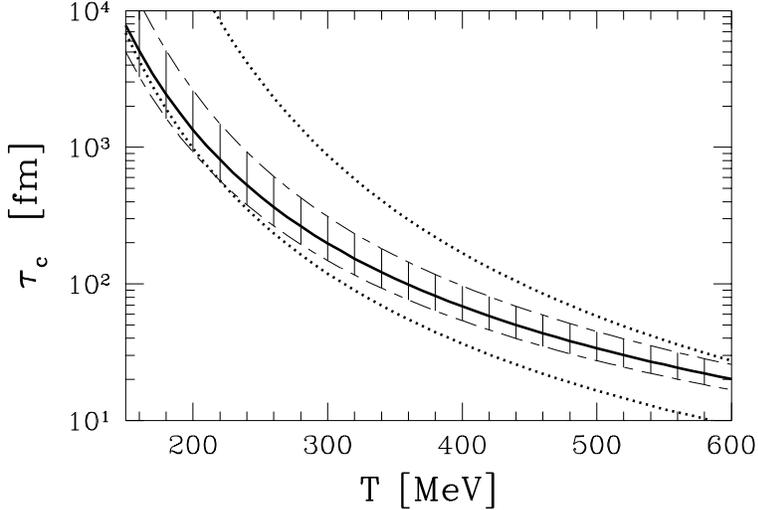}}
\vspace*{-0.3cm}
\caption{\small \label{figtaucc}
Solid lines: thermal charm relaxation constant in QGP, calculated
for running  $\alpha_s(M_{Z})=$ 0.115; 0.118, (indistinguishable),
$m_{\rm c}(M_Z)= 700$ MeV. Lower dotted line: for fixed $m_{\rm
c}=1.1$ GeV, $\alpha_{\rm s}=0.35$; upper doted  line: 
for fixed $m_{\rm c}=1.5$ GeV, $\alpha_{\rm s}=0.4$\,.
Hatched area: effect of variation $m_{\rm c}(M_Z)= 700\pm 50$~MeV.
}\end{figure}
We can now proceed to evaluate the relaxation times for
strangeness and charm production. The current calculation offers an
upper limit on the actual relaxation time, which may still be
smaller, considering that other mechanisms may contribute to
flavor production, as is shown by the dots in Eq.\,(\ref{qgpA}). 
We  show in Fig.\,\ref{figTaussrun}  also
the impact of a 20\% uncertainty in $m_{\rm s}(M_{{Z}})$, 
indicated by the hatched areas. This uncertainty is today
much larger compared to the uncertainty that arises from the
recently improved precision of the strong coupling constant
determination\,~\cite{Sch96}.  We note that
the calculations made at fixed values 
$\alpha_{\rm s}=0.5$ and $m_{\rm s}=200$~MeV 
(dotted line in Fig.\,\ref{figTaussrun}) are well within the band
of values related to the uncertainty in the strange quark mass.
 
Since charm is somewhat more massive compared to strangeness, there
is still less uncertainty arising in the extrapolation of the
coupling constant. Also the systematic uncertainty related to the
soft gluons (odd-$\alpha_{\rm s}$) terms are smaller, and thus the
relaxation times $\tau_{\rm c}$ we show in Fig.\,\ref{figtaucc}
are considerably better defined compared to $\tau_{\rm s}$. There
is also  less relative uncertainty in the value of charm mass.
We also show in Fig.\,\ref{figtaucc} (dotted lines) the fixed 
$m_{\rm c},\, \alpha_{\rm s}$ results with parameters selected to
border high and low $T$ limits of the results presented. It is
difficult to find a good comparative behavior of $\tau_{\rm c}$ 
using just one set of $m_{\rm c}$ and $\alpha_{\rm s}$. This may be
attributed to the importance of the mass of the charmed quarks,
considering that the threshold for charm production is well above
the average thermal collision energy, which results in emphasis of
the effect of running charm mass. In the high $T$-limit the choice
(upper doted line in Fig.\,\ref{figtaucc}) $m_{\rm c}=1.5$ GeV,
$\alpha_{\rm s}=0.4$ is appropriate, while to follow
the result at small $T$  (lower doted line in
Fig.\,\ref{figTaussrun}) we take a much smaller 
mass $m_{\rm c}=1.1$ GeV, $\alpha_{\rm s}=0.35$\,.

\subsubsection{Temporal evolution of the fireball}
In order to compute the production and evolution of
strangeness (and charm) flavor a more specific picture of the
temporal evolution of dense matter is needed. In a simple, qualitative
description, we assume that the hot, dense matter is homogeneous. 
We consider that, in Pb--Pb collisions at SPS,
the radial expansion is the dominant
factor for the evolution of the fireball properties such as
temperature/energy density and lifetime of the QGP phase. 
The expansion dynamics follows from two assumptions: \\ 
\noindent $\bullet$ the (radial) expansion 
is entropy conserving, thus the volume and temperature satisfy
\begin{equation}\label{adiaex}
V\cdot T^3=\,{\rm Const.}
\end{equation}
\noindent $\bullet$ the surface flow velocity is given by the 
sound velocity in  a relativistic gas
\begin{equation}
v_{\rm f}=1/\sqrt{3}\,.
\end{equation}
It is well understood that expansion into `vacuum' of relativistic
 matter can proceed at any velocity, up to velocity of light. However,
here $v_{\rm f}$ is understood not as the velocity of the first flow edge, 
it is the velocity of the flow of matter from interior of the fireball to
the near surface region, which cannot exceed the speed of sound in the 
matter, unless a shock is created.  We will return to this intricate 
issue in the near future~\cite{future}.

These two assumption imply the following explicit forms 
for the radius of the fireball and its average temperature:
 \begin{equation}
R=R_{\rm in}+{1\over \sqrt{3}}(t-t_{\rm in}),\
\label{T(t)}\qquad 
T={T_{\rm in}
\over{1+({t-t_{\rm in}})/\sqrt{3}R_{\rm in}}}.\nonumber
\end{equation}
We shall see below that if QGP formation is involved, 
a fit of strange antibaryons data either leads to direct emission
before expansion, or to emission from a surface expanding with 
just this velocity $v_{\rm f}$.

The initial conditions have been determined such that the energy per baryon is 
given by energy and baryon flow, and the total baryon number is 
$\eta (A_1+A_2)$, as stopped in the interaction region.
They are shown in table~\ref{initialc}.
 The radius are for zero impact parameter.
For this, equations of state of the 
QGP are needed, and we have employed our model~\cite{analyze2}  in which the
perturbative correction to the number of degrees of freedom were 
incorporated along with thermal particle masses.
 
\begin{table}[t]
\caption{\label{initialc}
The initial conditions for S--Pb/W at 200A GeV and Pb--Pb at 158A GeV
for different stopping values $\eta$.}
\begin{center} 
\begin{tabular}{l|ccccc} 
\hline\hline\vphantom{$\displaystyle\frac{E}{B}$}
&$ t_{\rm in}$ [\rm fm] & $\eta$ & $R_{\rm in}$  [\rm fm] 
& $T_{\rm in}$ [{\rm MeV}]& $\lambda_{\rm q}$\\
\hline
S--Pb/W &1&0.35&3.3&280&1.5\\
&1&0.50&3.7&280&1.5\\
\hline
Pb--Pb &1&0.50&4.5&320&1.6\\
&1&0.75&5.2&320&1.6

\end{tabular} 
\end{center} 
\vspace{-0.4cm}
\end{table} 

Allowing for dilution of the phase space density 
in expansion, we integrate~\cite{Rio95} a population 
equations describing the change in $\gamma_{\rm s}(t)$:
\begin{equation}\label{dgdtf}
\hspace*{-0.2cm}{{d\gamma_{\rm s}}\over{dt}}\!=\!
\left(\!\gamma_{\rm s}{{\dot T m_{\rm s}}\over T^2}
     {d\over{dx}}\ln x^2K_2(x)\!+\!
{1\over 2\tau_{\rm s}}\left[1-\gamma_{\rm s}^2\right]\!\right).
\end{equation}
Here K$_2$ is a Bessel function and $x=m_{\rm s}/T$. 
Note that even when $1-\gamma_{\rm s}^2<1$ we still can have 
a positive derivative of $\gamma_{\rm s}$, since the first term
on the right hand side of Eq.\,(\ref{dgdtf}) is always positive,
both $\dot T$ and $d/dx(x^2K_2)$ being always negative. This shows
that dilution due to expansion effects in principle can make the
value of $\gamma_{\rm s}$ rise above unity.
 
\begin{table}[t]
\caption{$\gamma_{\rm s}$ and  $N_{\rm s}/B$ 
in S--W at 200A GeV and Pb--Pb at 158A GeV
for different stopping values of baryonic number and energy 
$\eta_{\rm B}=\eta_{\rm E}$\,; computed for strange quark mass 
$m_{\rm s}(1 GeV)=200\pm40$ MeV, $\alpha_{\rm s}(M_Z)=0.118$\,.
\label{yield1}
}
\begin{center} 
\begin{tabular}{l|cc|cc} 
\hline\hline\vphantom{$\displaystyle\frac{E}{B}$}
$E_{\rm lab}$&\multicolumn{2}{|c}{S--W at 200A GeV}&%
\multicolumn{2}{|c}{Pb--Pb at 158A GeV}\\
\hline
$\eta_{\rm B}=\eta_{\rm E}$&0.35&0.5&0.5&0.75  \\
$\gamma_{\rm s}$& $0.53\pm 0.14$   &$0.65\pm 0.15$ &%
$0.69\pm 0.15$ &$0.76\pm 0.16$ \\
$N_{\rm s}/B$&$0.67\pm 0.16$&$0.70\pm 0.16$&%
$0.954\pm 0.20$&$0.950\pm 0.20$
\end{tabular} 
\end{center} 
\vspace{-0.4cm}
\end{table} 
Given  the relaxation constant $\tau_{\rm s}(T(t))$, these
equations can be integrated numerically, and we can obtain 
for the two currently explored experimental systems the values of the
two observables, $\gamma_{\rm s}$ and $N_{\rm s}/B$, which are given in 
table~\ref{yield1}. There is a considerable uncertainty due to 
the unknown mass of strange quarks. However, since this is 
 not a statistical but systematic 
uncertainty depending nearly alone on the value of the 
strange quark mass parameter, all the results presented will shift
together. We note further that there seems to be 
very little dependence on the stopping fractions in
the yield of strange quarks per baryon $N_{\rm s}/B$. Thus if 
the expected increase in  stopping is confirmed, we should
also expect a small increase by 15\% in specific strangeness yield.

\section{Hadronization from Fireball}\label{hadro}
\subsubsection{Fireball parameters}
We next introduce all the model parameters used in the
fit of the particle ratios, not all will be required in 
different discussions of the experimental data. For more details
about  the thermo-chemical parameters
we refer to the extensive discussion in the
earlier study of S--S and S--W data 
~\cite{acta96}. The key parameters are:\\ 
\indent 1) $T_{\rm f}$: the formation/emission/freeze-out temperature,
depending on the reaction model. 
$T_{\rm f}$ enters in the fit of abundance ratios of unlike particles
presented within a fixed $p_\bot$ interval. The temperature $T_{\rm f}$ 
can in first approximation be related 
to the observed high-$m_\bot$ slope $T_\bot$ by:
\begin{equation}\label{shiftT}
T_\bot \simeq T_{\rm f} \frac{1+v_\bot}{\sqrt{1-v_\bot^2-v_\parallel^2}}\,.
\end{equation}
In the central rapidity region the longitudinal 
flow $v_\parallel\simeq 0$, in order to assure
 symmetry between projectile and target. Thus as long as  $T_{\rm f}<T_\bot$, 
we shall use Eq.\,(\ref{shiftT}) setting $v_\parallel=0$ 
to estimate  the transverse 
flow velocity $v_\bot$ of the source.\\ 
\indent 2) $\lambda_{\rm q}$: the light quark fugacity.
We initially used in our fits both $u,\,d$-flavor fugacities
$\lambda_{\rm u}$ and $\lambda_{\rm d}$, but
we saw that the results were equally adequate
without allowing for up-down quark asymmetry, using  the 
geometric average $\lambda_{\rm q}=\sqrt{\lambda_{\rm u}\lambda_{\rm d}}$; 
moreover the 
fitted up-down quark fugacity asymmetry was found
 as expected in 
our earlier analytical studies~\cite{analyze1}.\\ 
\indent 3) $\lambda_{\rm s}$:  the strange quark fugacity. A source in which
the carriers of $s$ and $\bar s$ quarks are symmetric this parameter
should have the value $\lambda_{\rm s}\simeq1$, in general in a re-equilibrated 
hadronic matter the value of $\lambda_{\rm s}$ can be determined requiring
strangeness conservation.\\  
\indent 4) $\gamma_{\rm s}$:  the strange phase space occupancy. Due to 
rapid evolution of dense hadronic matter it is in general highly 
unlikely that the total abundance of strangeness can follow the 
rapid change in the conditions of the source, and thus in general 
the phase space will not be showing an overall abundance equilibrium 
corresponding to the momentary conditions.\\  
\indent 5)  $R_{\rm C}^{\rm s}$: in table~\ref{t1} we note this
parameter describing the relative 
off-equilibrium abundance of
strange mesons and baryons, using thermal equilibrium abundance as reference.
 
Parameter $R_{\rm C}^{\rm s}$ is needed, when we have constraint 
on the strangeness
abundance and/or when  we address the abundance of mesons 
since there is no a priori assurance that the relative
production/emission strength of strange mesons and baryons should 
proceed according to 
relative strength expected from thermal equilibrium. Moreover, it is 
obvious that even if re-equilibration of particles in hadronic gas should occur, 
this parameter will not easily find its chemical equilibrium 
value $R_{\rm C}^{\rm s}=1$
as we alluded to in section 1. However, 
due to reactions connecting strange with non-strange particles we 
expect $R_{\rm C}^{\rm s}=R_{\rm C}$, where $R_{\rm C}$ is the same ratio  for non-strange mesons 
and baryons, using thermal abundance as reference. The value of $R_{\rm C}>1$
implies meson excess abundance per baryon, and thus excess specific
entropy production, also expected in presence of
 color  deconfinement~\cite{entropy}. 

The relative number of particles of same type 
emitted at a given instance by a hot source is obtained by
noting that the probability to find all the $j$-components contained within 
the $i$-th  emitted particle is
\begin{equation}\label{abund}
N_i\propto \gamma_{\rm s}^k\prod_{j\in i}\lambda_je^{-E_j/T}\,,
\end{equation}
and we note that the total energy and fugacity of the particle is:
\begin{equation}
E_i=\sum_{j\in i}E_j,\qquad \lambda_i=\prod_{j\in i}\lambda_j\,.
\end{equation}
The strangeness occupancy $\gamma_{\rm s}$ enters Eq.\,(\ref{abund})
with power $k$,  which equals the number of strange and anti-strange quarks in 
the hadron $i$. 
With $E_i=\sqrt{m_i^2+p^2}=\sqrt{m_i^2+p_\bot^2}\cosh y $ 
 we integrate over the transverse momentum range 
as given by the experiment (here $p_\bot>0.6 $ GeV)
taking  central rapidity region $y\simeq 0$
to obtain the relative strengths of particles produced. 
We then allow all hadronic resonances to disintegrate 
in order to obtain the final relative multiplicity of `stable' particles
required to form the observed particle ratios. 
This approach allows to compute the relative
strengths of strange (anti)baryons both in case of 
surface emission and equilibrium disintegration of a particle gas since
the phase space occupancies are in both cases properly accounted for by 
Eq.\,(\ref{abund}). The transverse flow phenomena enter in a similar fashion into
particles of comparable mass and are not expected to
influence particle ratios. 
 Finally we note  that  particles which are easily influenced by the
medium, such as $\phi$, require a greater effort than this
simple model, and are also not explored in depth here.

{\begin{table}[t]
\caption{
Values of fitted statistical parameters within thermal model,
for 158A GeV Pb--Pb strange particle production data. 
Superscript star `*': a fixed input value for equilibrium 
hadronic gas;  subscript `$|c$':
value is result of the imposed strangeness conservation constraint. 
$\chi^2$ is  the total relative square 
error of the fit for  all data points used. First result line: direct emission
QGP model, no meson to baryon ratio fit. Second line: same, but with strangeness
conservation yielding $\lambda_{\rm s}$, and $R_{\rm C}^{\rm s}$ variable. 
Line three: as in line two, in addition the meson to baryon 
ratio Eq.\,(\protect\ref{klratio}) is fitted. Line four: hadronic gas fit
including the ratio Eq.\,(\protect\ref{klratio}).
\label{t1}
}
\vspace{0.1cm}
\begin{center}
\begin{tabular}{l|ccccc} \hline\hline 
$T_{\rm f} [MeV]$& $\lambda_{\rm q}$&$\lambda_{\rm s}$&
$\gamma_{\rm s}$&$R_{\rm C}^{\rm s}$& $\chi^2$$\vphantom{\ds\frac{\Xi}{\Lambda}}$ \\
\hline
                   272 $\pm$ 74
                 & 1.50 $\pm$ 0.07
                 &   1.14 $\pm$ 0.04
                 &   0.63 $\pm$ 0.10
                 &   ---
                 &   1.0  \\

                   272 $\pm$ 74
                 & 1.50 $\pm$ 0.08
                 &   1.14$_{|c}$
                 &   0.63 $\pm$ 0.10
                 &   4.21$\pm$ 1.88
                 &   1.0  \\
\hline

                     151 $\pm$ 10
                 &   1.54 $\pm$ 0.08
                 &   1.13$_{|c}$
                 &   0.91$\pm$0.09
                 &   0.85$\pm$0.22
                 &     1.5      \\

                     155 $\pm$ 7\phantom{0}
                &  1.56 $\pm$ 0.09
                &  1.14$_{|c}$
                &  1$^*$
                &  1$^*$
                &   7.6   \\

\end{tabular}\\
\end{center} 
\vspace{-0.4cm} 
\end{table}}


We obtain the least square fit for the eight above reported
 (anti)baryon ratios. 
Our first approach is motivated by the reaction picture consisting of
 direct emission from the 
QGP deconfined fireball. The value of statistical parameters controlling
the abundances are thus free of constraints arising in an equilibrated
hadronic gas (HG) state~\cite{analyze1}. 
The fitted thermal parameters are presented in the
first line of table~\ref{t1} along with the total $\chi^2$ for
the eight  ratios. The fit is quite good, the 
error shown corresponds to the total accumulated error from 8 measurements;
even if one argues 
that it involves 4 parameters to describe 5 truly independent quantities, 
the statistical significance is considerable, considering that 8 different 
measurements are included.
Such a free fit does not know that we are expecting that the final 
state comprises a balance $\langle s-\bar s\rangle=0$.
In order to estimate what would be implied by strangeness conservation 
constraint among emitted hadrons we present in second line of table~\ref{t1} 
the result of a fit assuming that the value of $\lambda_{\rm s}$ 
is result of the conservation constraint $\langle s-\bar s\rangle= 0$,
and allowing  $R_{\rm C}^{\rm s}\ne 1$, for there should 
be no chemical equilibrium 
among the emitted  strange mesons and strange  baryons
in a sudden and early QGP disintegration. The statistical  error is found
 the same as in line 1, since this approach substitutes one parameter
 ($\lambda_{\rm s}$) by another ($R_{\rm C}^{\rm s}$). 
The implication of this fit is that 
there must be either an excess of strange mesons or 
depletion of strange baryons
compared to thermal equilibrium expectation, since as we recall 
$R_{\rm C}^{\rm s}=C_{\rm M}^{\rm s}/C_{\rm B}^{\rm s}$ with $C_i$ 
being the yield of particles, normalized
to one for thermal equilibrium yield. 
In any case we see that strange meson excess is required, which
is consistent with excess of entropy production.

\begin{figure}[tb]
\vspace*{-1.6cm}
\centerline{\hspace*{-.8cm}
\psfig{width=8cm,figure=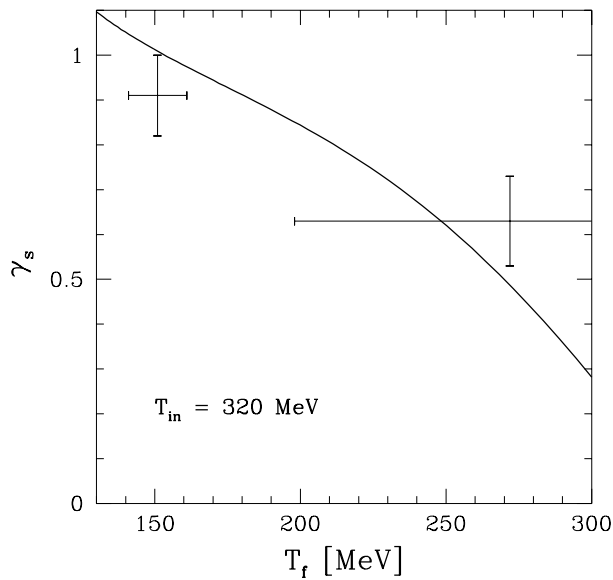}
}
\vspace*{-1.3cm}
\caption{
QGP strangessnes occupancy $\gamma_{\rm s}$ as function
of temperature $T_{\rm f}$ at time of particle production, for 
intial temperature $T_{\rm in}=320$ MeV, with 
$\gamma_{\rm s}(T_{\rm in})=0.1$. 
\label{gammTf}
}

\end{figure}

The errors seen in the two first lines of table~\ref{t1} 
 on the statistical parameters arise in part from strong
correlations among them. In particular
  the very large error in $T_{\rm f}$ arise from strong
(anti) correlation with $\gamma_{\rm s}$: we find that a
change in one of these parameters is associated with a change in the
other at 80\% level.  Further 
information about the relation of $T_{\rm f}$ and $\gamma_{\rm s}$ 
may be garnered from theoretical considerations. We 
evaluate using our  dynamical strangeness production
model in QGP  how the
value of $\gamma_{\rm s}$ depends on the temperature of particle 
production $T_{\rm f}$. The most important parameter in such 
a theoretical  evaluation is the initial temperature at which the deconfined
phase is created. As noted above,  we estimated this temperature 
at $T_{\rm in}=320$ MeV~\cite{analyze2}.
Further uncertainty of the calculation arises from the strange quark mass 
taken  here  to be $m_{\rm s}(1\mbox{\,GeV})=200$ MeV. We recall that 
the strength of the 
production rate is now sufficiently constrained by the measurement of 
$\alpha_{\rm s}(M_Z)$.
We choose  a geometric size which comprises a baryon number $B= 300$
 at $\lambda_{\rm q}\simeq 1.5$, and have verified that
 our result will be little dependent on small
variations in $B$. We show in Fig.\,\ref{gammTf} how the computed
$\gamma_{\rm s}$ depends
on formation temperature $T_{\rm f}$. The cross to the right
shows our  fitted value from line 1 or 2 in table~\ref{t1}. 
It is gratifying to see that it is consistent with the 
theoretical expectation  for early formation of the strange 
(anti)baryons. The relative smallness of $\gamma_{\rm s}$,
despite the hight strangeness yield, is 
clearly related to the high temperature of particle production.
We will below discuss the opposite scenario: the late freeze-out of 
strange particles.

The relatively high value of temperature $T_{\rm f}$ we obtained in the 
reaction picture with primordial particle emission is consistent
with the experimental inverse slopes of the spectra.
We are thus led to the conclusion that 
 as far as the fitted temperatures and slopes are 
concerned, it is possible that  the high $m_\bot$ 
strange (anti)baryons we have described could have been emitted directly 
from a primordial (deconfined) phase before it evolves into final 
state hadrons.

We now consider at the experimental inverse $m_\bot$  slopes.
In the common $p_\bot$ range of WA97 and NA49 experiments the transverse 
mass spectrum of $\Lambda$ and $\overline\Lambda$ obtained by NA49 is
very well describe by an exponential~\cite{NA49B}. 
A thermal model motivated fit of the inverse slope 
(temperature) yields $T_\bot^\Lambda=284\pm15$ MeV and 
$T_\bot^{\overline{\Lambda}}=282\pm20$ MeV. This is consistent with
the mid-rapidity proton and antiproton slope of the NA44 experiment: 
$T_p=289\pm7$ MeV and $T_{\overline{p}}=278\pm9$ MeV. For $\Xi+\overline{\Xi}$
a consistent value $T_\Xi=290$ MeV is also quoted by the NA49--collaboration~~\cite{NA49O}. 
We note that because the baryon masses are large, all 
these slopes are at relatively
high $m_\bot>1.3 $ GeV (for nucleons, in NA44, $m_\bot>1 $ GeV). 
Systematically smaller inverse-transverse slopes    
are reported at smaller $m_\bot$, for kaons $T_\bot^K\simeq 213$--$224$ MeV 
for $0.7 <m_\bot<1.6$ GeV in NA49~\cite{NA49B} and 
$T_\bot^{K^+}= 234\pm6,\,T_\bot^{K^-}=235\pm7$ MeV in NA44~\cite{NA44};
and 155--185 MeV for $\pi$,~\cite{NA49B,NA44},  depending 
on the range of $p_\bot$, but here we have to remember that pions
are known to arise primarily from resonance decays. 
An increase of $T$ with $m_\bot$ 
 is most naturally associated with the effects of transverse flow of
the source.

Is QGP primordial emission hypothesis also consistent with the chemical fugacities  we have obtained? The chemical condition is 
fixed to about 5\% precision, and there is 40\% anti-correlation
between the two fugacities $\lambda_{\rm q}$ and $\lambda_{\rm s}$. The
information that
$\lambda_{\rm s}\ne1$ is contained in at least two particle
abundances; arbitrary manipulation of the reported yields of one particle 
abundance did not reduce the value $\lambda_{\rm s}$ to unity. 
Since $\lambda_{\rm s}\ne 1$ 
by  4\,s.d.  it is highly unlikely that $\lambda_{\rm s}=1$ 
is found after more data is studied. While one naively expects 
$\lambda_{\rm s}^{\rm QGP}=1$, to assure the strangeness balance 
$\langle s-\bar s\rangle=0$, there must be 
a small deviations from this value, even if the emitted particles were to
reach asymptotic distances without any further interactions: 
in presence of baryon density the deconfined state is not
fully symmetric under interchange of particles with antiparticles. 
A possible mechanism to distinguish the strange and anti-strange quarks 
arises akin to the effect considered for the $K^-/K^+$ asymmetry  in baryonic 
matter~\cite{Shu92,Go92}: there is asymmetric scattering  strength on 
baryon density $\nu_{\rm b}$
 which causes presence of a mean effective vector potential
$W$. Similarly, strange quark interaction with baryon density would lead to a
dispersion relation 
\begin{equation}\label{Ws}
E_{\rm s/\bar s}=\sqrt{m_{\rm s}^2+p^2}\pm W \, ,
\end{equation}
and this requires in the statistical approach that the 
Fermi distribution for strange and anti-strange 
quarks acquires a  compensating fugacity 
$\lambda_{{\rm s},\bar {\rm s}}=e^{\pm W/T}$ 
to assure strangeness balance in the deconfined phase. In linear
response approach $W\propto\nu_{\rm b}$  consistent with both $W$  and
baryon density $\nu_{\rm b}=(n_{\rm q}-n_{\bar {\rm q}})/3$, 
being fourth component of a Lorentz-vector. 
It is clear for intuitive reasons, as well as given experimental 
observations, that the baryon stopping and thus density increases
considerably comparing the S and Pb induced reactions in the energy domain
here considered. We also recall that in S--W reactions 
 $\lambda_{\rm s}^{\rm S}\simeq 1.03\pm 0.05$~\cite{analyze}. 
Should in the dense matter fireball the 
baryon density $\nu_{\rm b}$ grow by factor 2--4 
as the projectile changes from S to Pb, this alone would
consistently explain  the appearance of the value $\lambda_{\rm s}=1.14\pm0.04$
obtain using $W\propto\nu_{\rm b}$ scaling. 
It is worth noting that the value $W\simeq 38$\,MeV suffices here. Note also
that the Coulomb potential effect on the charge of the strange quarks is
of opposite magnitude and about 1/5--1/6 of the here required strength.

\subsubsection{Late emission scenario: HG with or without QGP?}
A generally favored picture of particle production involves flow expansion 
of the primordial phase till a transition temperature is reached, 
at which time the final state hadrons are produced, and 
soon thereafter freeze out. These particles may
directly reach a detector or re-equilibrate and appear to the
observer as if emitted from a HG phase, except that 
entropy/strangeness  excess effect should remain. 
In order to force our data fit to converge to such a late particle 
production scenario we introduce the experimental result,
which was essential for such an argument in the S--Pb induced 
reactions. The quantity of interest is ratio of particle yields 
between particles of very different mass. Thus
in lines 3, 4 in table~\ref{t1} we include in the fit also as experimental
input an estimate of the hyperon to kaon ratio, thus 
altogether we now fit 9 data points.

We note that the NA49 spectra
~\cite{NA49B} of kaons and hyperons have a slightly overlapping domain of 
$m_\bot$. We recall that the slopes are 
not exactly equal, thus all we can do is to
try to combine the two  shapes, assuming continuity consistent with flow, 
and to estimate the relative normalization of both that would place
all experimental points on a common curve. We have carried out this procedure 
and obtained:
\begin{eqnarray}\label{klratio}
\left.\frac{\Lambda}{K^0_{\rm s}}\right|_{m_\bot}\simeq 6.2\pm1.5\,.
\end{eqnarray}
Note that there is a tacit presumption in our approach 
that a similar effective  $\Delta y$ interval
was used in both spectra.  We recall that this ratio
was $4.5\pm0.2$ in the S-W data~\cite{WA85K}.

Our approach in the third line corresponds to a freeze-out from
`cold' QGP phase, in that we allow the 
abundance parameters $\gamma_{\rm s}$ and $R_{\rm C}^{\rm s}$ 
to deviate from HG equilibrium values. 
We note that this cold-QGP alternative has a 
very comparable statistical significance as the hot-QGP.
Given the low temperature and high  $m_\bot$ 
inverse slopes we must have considerable transverse flow.
The  computed flow velocity at freeze-out is $v_{\rm f}=0.51$\,. 
This is just below the  relativistic sound velocity 
$v_{\rm s}=1/\sqrt{3}=0.58$\,, which we have assumed.
 In  Fig.\,\ref{gammTf} the cross to the left shows the 
result of the fit we just described; allowing for potentially smaller expansion 
velocity and all the above discussed uncertainties in the 
computation, this result must also be seen as a very good agreement 
between the result of data fit and the theoretical calculation.
This also means that we cannot distinguish in the present data between 
early formation of strange antibaryons and an expansion model followed
by direct global hadronization. 

Since the fitted values of $\gamma_{\rm s}$ and $R_{\rm C}^{\rm s}$ allow the HG 
equilibrium, we attempt such a fit in line 4, where the particle
yields are fitted constrained for HG equilibrium, and
we use strangeness conservation to evaluate the strangeness fugacity 
$\lambda_{\rm s}$. We show the result of the fit in the last line 
of  table~\ref{t1} and in particular we note:
\begin{eqnarray}
T_{\rm f}&=155\pm 7\,\mbox{MeV},\ &\rightarrow v_\bot
\simeq 0.5\simeq v_{\rm s};\nonumber\\
\lambda_{\rm q}&=1.56\pm0.09,\hspace*{0.25cm} \ &\rightarrow\  \lambda_{\rm s}=1.14;\nonumber\\
\chi^2/9&=0.84,\hspace*{1.4cm} \ &\rightarrow \ \ \mbox{C.L.}>60\%\,.
\end{eqnarray}
We recall that the baryo-chemical potential is given in 
terms of $T$ and $\lambda_{\rm q}$, specifically 
$\mu_{\rm b}=3T\ln \lambda_{\rm q}$, and we 
find $\mu_{\rm b}=204\pm10$ MeV in this hadronic gas condition. 
We note that while the quality of the fit has degraded, it 
still has considerable statistical significance. 
It would appear to be a `good' fit to the naked, an unequipped eye. 

\begin{figure}[tb]
\vspace*{-1.6cm}
\centerline{\hspace*{-.8cm}
\psfig{width=8cm,figure=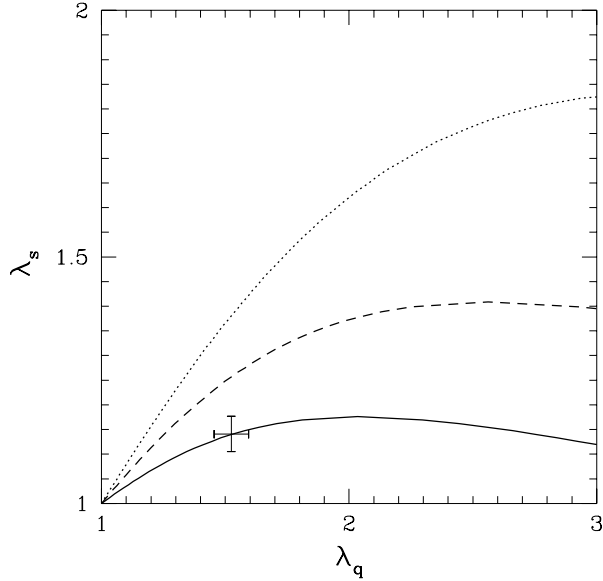}
}
\vspace*{-1.4cm}
\caption{Strangeness conservation constraint in HG
as function of freeze-out $\lambda_{\rm q}$: the 
lines correspond to different freeze-out temperatures 
$T_{\rm f}$ (solid 160 MeV, dashed 140 MeV and dotted 120 MeV).
The cross corresponds to
the chemical freeze-out we determined above.
\label{s-sbarHG}
}

\end{figure}

Unlike S-induced reactions, the  hypothesis of an equilibrated HG 
in the final state cannot be easily 
discarded in Pb--Pb collisions since the 
thermo-chemical parameters are extraordinarily consistent 
with  this hypothesis and the principle of strangeness conservation, as we 
illustrate in
Fig.\,\ref{s-sbarHG}. Here, the cross corresponds to the fitted properties 
of the particle source, while the lines correspond to the constraint of the
HG gas source to yield $\langle s-\bar s\rangle=0$ at finite baryon 
density represented by the value of $\lambda_{\rm q}$. Thus the strange quark 
fugacity is in general not zero and the cross falls just 
on freeze-out at $T=160$ MeV when the meson-baryon abundance is equilibrated.
The slight difference in freeze-out value to results shown in last line of 
table~\ref{t1} is result of using the WA97 
$\bar\Lambda/\Lambda$ value Eq.\,(\ref{rat1}) 
rather than the averred value Eq.\,(\ref{rat2}).

If HG is indeed present in the final state, 
 the proper interpretation of these data, and the likely reaction
scenario, compatible with our earlier work on S induced reactions 
~\cite{acta96} is as follows; the relatively large fireball of 
dense and deconfined matter disintegrates and produces 
dense, confined hadronic gas in which strange particles have time to 
re-scatter and to establish relative chemical equilibrium. 
A possible test of this hypothesis would be to see variation of 
the chemical parameters  as the size of the fireball changes 
with impact parameter (centrality of collision) 
since re-equilibration should diminish for small reaction volume.
However, such data are presently not available, and 
there is no indication that indeed a change of the strange (anti)baryon 
yields occurs as the centrality of the interaction is reduced. 
On the other hand, the specific entropy and strangeness should
comprise a signal of some new physics should
formation and expansion of QGP phase, followed by 
re-equilibration into HG phase, and freeze-out have occurred.
We will now consider the magnitude of these effects:\\

\noindent{\bf Strangeness re-equilibration}\\
When  HG emerges from initial dense  QGP phase, the
number of strange quark pairs does not change, but the 
phase space density of strangeness changes, since the phases
are different. Because the HG phase has generally a smaller
phase space density of strangeness than QGP, to conserve 
strangeness, there will be a jump in the phase space occupancy 
$\gamma_{\rm s}$ during the transformation of QGP into HG, as there 
is a jump in the strange quark fugacity. The important point is 
that this could lead to significantly 
overpopulated  HG phase ($\gamma_{\rm s}>1$). This phenomenon 
can be easily quantified as follows:
the observed value of $\gamma_{\rm s}^{\rm HG}$ is related to the pre-phase
change value  $\gamma_{\rm s}^{\rm QGP}$ by introducing the enhancement 
factor we wish to determine:
\begin{equation} \label{F1}
\gamma_{\rm s}^{\rm HG}\equiv F_\gamma \gamma_{\rm s}^{\rm QGP}
\end{equation}
A simple way to compute the value of the saturation enhancement 
factor $F_\gamma$ is to study the
abundance of strangeness per baryon number 
before and after phase transition.
\begin{equation}\label{F2}
F_\gamma =\frac{s/b|_{QGP}}{s/b|_{HG}} 
    = \frac{\gamma_{\rm s}^{\rm QGP}}{\gamma_{\rm s}^{\rm HG}} 
                    f(T_{\rm f},\lambda_{\rm q},\gamma_{\rm s}^{\rm HG})\,.
\end{equation}
The last expression arises as follows: 
on the QGP side the abundance of 
strangeness is to a good approximation proportional 
to $\gamma_{\rm s}^{\rm QGP}$ and is the 
integral of the strange quark phase 
space, we evaluate it assuming that 
$m_{\rm s}/T_{\rm QGP}\simeq 1$. There is no
dependence on chemical properties of 
the plasma. On HG side, at freeze-out 
we have to evaluate the strangeness abundance 
from the strange particle 
partition function given in Eq.\,(16) of~\cite{analyze1}, 
supplemented by the
now relevant term comprising
 $s\bar s$--$\eta,\ \eta',\ \phi$ states
and their resonances. The sum includes 
a terms proportional to 
$(\gamma_{\rm s}^{\rm HG})^n$, with n=1, 2, 3, 
indicating strangeness content of hadrons. 
The leading kaon and hyperon term is proportional 
to $\gamma_{\rm s}^{\rm HG}$ and hence
we have above result, Eq.\,(\ref{F2}). 
We thus obtain, combining Eqs.\,(\ref{F1}) and Eq.\,(\ref{F2}),
\begin{equation}\label{F3}
F_\gamma^2=F_\gamma
    \frac{\gamma_{\rm s}^{\rm HG\ }}{\gamma_{\rm s}^{\rm QGP}}=
\left.\frac{s}{\gamma_{\rm s}b}\right|_{QGP}\cdot \left.\frac{\gamma_{\rm s} b}{s}\right|_{HG}\,, 
\end{equation}
where the right hand side now compares the properties of the two 
phases at the boundary between them and we can evaluate it using 
the theoretical equations of state. In analyzes of an 
experiment we would take the  freeze-out parameters 
determined by the fit to data.

\begin{figure}[tb]
\vspace*{-1.6cm}
\centerline{\hspace*{-.8cm}
\psfig{width=8cm,figure=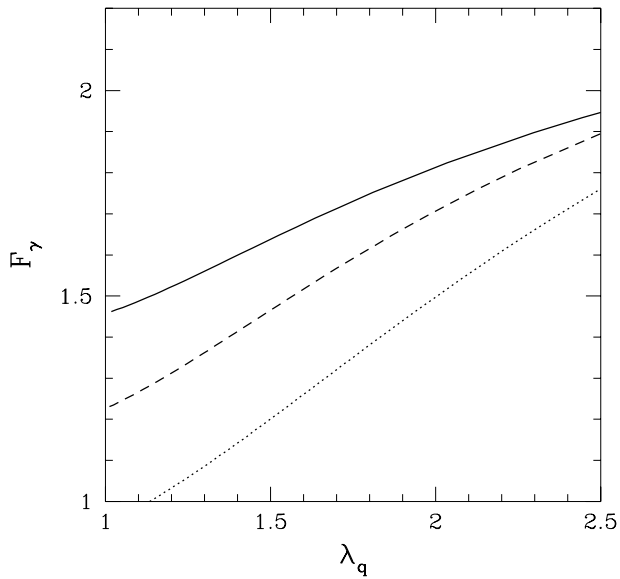}
}
\vspace*{-1.4cm}
\caption{Strangeness phase space enhancement factor as function of
HG freeze out $\lambda_{\rm q}$ for $T_{\rm f}=160$ (solid line),
$T_{\rm f}=140$ (dashed line), $T_{\rm f}=120$ (dotted line). Computed
for $\gamma_{\rm s}^{\rm HG}=1$, $m_{\rm s}/T_{\rm QGP}=1$, including in HG phase
kaons, hyperons, cascades, $\eta$, $\phi,\, \Omega$, and imposing 
strangeness conservation constraint to determine~$\lambda_{\rm s}$. 
\label{fgamma}
}

\end{figure}
We show, 
in Fig.\,\ref{fgamma},  the strangeness enhancement factor as function of 
$\lambda_{\rm q}$ for several freeze-out 
temperatures $T_{\rm f}=160,\,140,\,120$ MeV,
with $\lambda_{\rm s}$ fixed by strangeness conservation constraint. We see 
that $F_\gamma$ varies  typically between 1.5 and 2, and is specifically 
1.6 for the parameter rage  of Pb--Pb collisions here discussed. 
This means that observing the value
$\gamma_{\rm s}\simeq 1$ really means an underlying 
value $\gamma_{\rm s}^{\rm QGP}\simeq 0.6$.
Conversely, should we be able to create a  longer lived or hotter QGP state 
we could expect to observe in the HG phase $\gamma_{\rm s}^{\rm HG}$ as large as
1.5--2. Such over-saturation of the phase space would 
be a rather strong smoking gun
pointing to the formation of the QGP phase.\\ 

\noindent{\bf Entropy and particle excess}\\
Another way to argue for the formation of QGP  in early stages of
an expansion  scenario of the fireball is to measure 
 the specific entropy experimentally,
for example by measuring the quantity
\begin{equation}    
   D_{\rm Q} \equiv \left({\displaystyle{dN^+\over dy} - {dN^- \over dy}}
\right) \left/\left( {\displaystyle{dN^+\over dy} + {dN^- \over dy}}\right) \,,
\right.\end{equation}
which we have shown to be a good measure of the entropy content~\cite{analyze}.
We note that in the numerator of $D_{\rm Q} $ the charge of particle
pairs produced cancels and hence this value is effectively a measure of
the baryon number, but there is a significant correction arising from
the presence of strange particles. The denominator is a measure of the
total multiplicity --- its value is different before or after disintegration
of the produced unstable hadronic resonances. Using as input the
distribution of final state particles as generated within the hadron gas
final state it is found~\cite{entropy} that $D_Q\cdot S/B$  is nearly
independent of the thermal  parameters and varies between 4.8, before
disintegration of the  resonances,  to 3 after disintegration. 
\begin{figure}[tb]
\vspace*{-1.6cm}
\centerline{\hspace*{-.8cm}
\psfig{width=8cm,figure=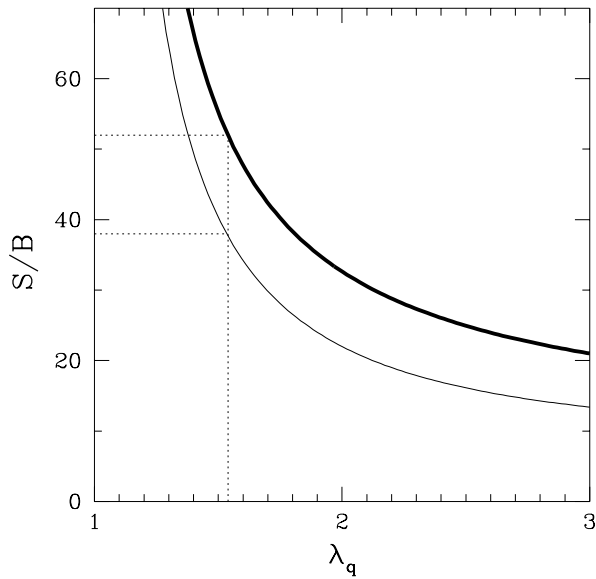}
}
\vspace*{-1.4cm}
\caption{
QGP (thick line) and HG (thin line, $T=155$ MeV) entropy per baryon 
$S/B$ as function of light quark fugacity $\lambda_{\rm q}$. Dotted lines
guide the eye for the here interesting values.
\label{entro1}
}
\end{figure}
 
To obtain a measure of the particle
excess, we show in  Fig.\,\ref{entro1} the specific entropy  per baryon
$S/B$ content in  dense hadronic matter  as a function of  light 
quark fugacity $\lambda_{\rm q}$. The thick line addresses
the deconfined QGP  phase, the thin line the confined HG phase at $T=155$ MeV, 
with the strange quark fugacity $\lambda_{\rm s}$ 
being determined from the strangeness conservation condition. While the 
QGP result is largely independent of temperature (only other,
aside of $T$ dimensioned quantity, is $m_{\rm s}$), the HG result involves 
the values of all hadron masses and hence is
dependent on $T$. The 12 units of entropy 
difference between the two phases for the here interesting range of 
fugacity $\lambda_{\rm q}=1.5$--$1.6$ implies that we should expect an excess of 
about 3 mesons per baryon if the deconfined phase is formed.
We compare with a HG at $T=155$ MeV; 
should the HG phase of interest be hotter, this 
difference between QGP and HG grows, since the baryon density in HG grows 
much faster than entropy for the baryon mass is well below the temperature
range and thus a change in the factor $m_{\rm N}/T$ matters, while the change 
in $m_\pi/T$ is immaterial. In other work HG phase at 190 MeV is often 
considered, and there the difference between QGP and HG properties turns out 
to be as large as factor two~\cite{entropy}. 

\section{Discussion and Outlook: QGP} \label{disc}
\subsubsection*{Vacuum and the QCD energy scale}
We have relegated to the vacuum properties all scales related
to hadronic structure. But where-from are these scales derived?
It seems that at present there is indeed no `low energy' practical 
understanding  of the hadronic  scale. While ordinary QED matter 
scale is governed by the atomic Bohr radius
 $a_0=\alpha_e^{-1}/m_e$, there is in QCD no obvious
relation between quark masses and hadronic spectra and sizes.
The scale of the vacuum condensates (see Section \ref{BagM}), 
and thus of the vacuum  condensation energy which results e.g. from the 
Lattice-QCD approach~\cite{lattice,EdGM98} is a measured property of
the physical ground state, and does not relate directly to any fundamental 
property of the QCD action alone; the  $\Lambda_0$-parameter of the strong 
interaction coupling constant $\alpha_s$ (see Eq.\,(\ref{Lambdarun}) and 
the middle-section of Fig.\,\ref{fig-a1}) 
has no further meaning, since in fact the  strength of  this coupling 
at our energy scale derives from  an initial value problem: for a given measured 
value of $\alpha_s$ at some energy scale, we integrate the renormalization 
group equations  towards the low energy domain (see Section \ref{sprod})
to obtain the needed interaction strength, and the definition of 
$\Lambda_0$-parameter than arises in terms of the initial values.  

It is not the $\Lambda_0$-parameter in the strong coupling constant $\alpha_s$ but
the initial point, thus e.g. the grand unification scale, which is defining for 
us the hadronic energy scale. It is quite possible, though not obvious at all,
that $\Lambda_0$  establishes  the strength of vacuum
fluctuations, and thus one would be tempted to conclude that hadronic mass scales
are the result of a `dimensional transmutation' mechanism which 
transports and reduces the grand unification scale to our physical domain.
This conclusion is not dependent on the scheme of the 
unification model employed, so the reader who is not happy with initial 
value $\alpha_s=\alpha_e$ and the associated huge scales $10^{16}$--$10^{18}$ GeV
can equally well use as the input-scale to strong interactions any other
initial point, as long as the result yields the now well established experimental
value~\cite{Sch96}  $\alpha_s(M_{\rm Z}\simeq 0.118\pm 0.03$. It is important
here to stress that there is no scenario that would yield the 
light quark mass as a significant scale, since the scale properties of the 
physical vacuum always dominate. There is moreover no 
obvious means to use the masses $m_s,m_c$ which would be 
easy to associate with QCD  condensate scales, since in such an 
approach there would be no understanding why just two `medium' and not 
the other quarks determine the scales, nor is there a noticeable regularity among
the mass parameters of matter (quarks, leptons) at any scale~\cite{FK98}. 

\subsubsection{Nuclear relativistic collisions and strangeness}
Considering these  strong and unexpected conclusions about the 
fundamental nature of strong (nuclear) interactions and its  far reaching 
implications, we must assure that the foundations on which these 
arguments are build are exceptionally solid. The unique 
tool to study the QCD vacuum is generally accepted to be the 
process of its change, the `micro' bang in the laboratory. 
In the relativistic heavy ion collision 
experiments we seek to prove by finding a way of freeing the 
color  charge of quarks that the effect of confinement arises
from quantum fluctuations in the QCD vacuum 
structure.  We hope and expect to produce in these 
collisions regions of space filled with mobile color charges of
quarks and gluons, forming `QGP', the 5th phase of matter comprising
a Quark-Gluon (color charge deconfined) Plasma. 

Strange particle production is today appreciated as one of the
most interesting hadronic observables  of dense, strongly interacting
matter and much of the current theoretical and experimental effort in study
of relativistic nuclear collisions is devoted to  this topic.
Our work  concentrates on the exploration of
the pattern of production and evolution of hadronic particles carrying
strangeness flavor. Our continued interest in the subject  arises from
the realization that the experimentally observed anomalous production of 
(strange) antibaryons cannot be
interpreted without introduction of some new physical phenomena, and the
hope and expectation is that when more systematic experimental data is 
available, we will be able to make the argument for deconfinement which
will be generally accepted. 

In our work we assume that thermal quark-gluon degrees
of freedom are at the basis of many of the hadronic particle production
phenomena in relativistic hadron reactions. 
Many simple, but subtle experimental observations point in this 
natural direction. For example, many of the measured $m_\bot$ spectra 
in S-- and Pb--induced reactions have the same shape for strange
baryons and antibaryons of the same kind, and even  different 
kinds of particles when considered in same range of $m_\bot$ show 
the same inverse transverse slope (=spectral temperature). 

In passing we note that there has been much discussion
of the fact that depending on particle mass, the slopes of the 
particle spectra show different temperatures. This behavior is possibly
arising from the fact that for different particles different ranges of 
$m_\bot$ were considered. This is thus evidence for 
significant expansion flow phenomena governing final stages of QGP 
disintegration and in no way does this behavior of the spectra  contradict 
the picture of local thermal equilibrium.

We have recently  improved the two particle collision 
cross sections leading to strangeness 
and charm production. This was done using the recent advances in  
measurement of the strong coupling constant $\alpha_{\rm s}$
at the scale of $M_{{Z}}$, a result which eliminated the need for 
the introduction of an arbitrary coupling constant in flavor production
calculations. Using simple dynamical models of dense matter 
evolution we are able to obtain the final state abundance and phase
space occupancy of strangeness originating in the deconfined phase.
From this condition at the point that deconfined phase ceases to 
exist and hadronizes, we develop using conservation laws, constrains on
the hadronic particle abundances. 

Relative abundances of strange hadrons allowed us to investigate
the chemical equilibrium. We find that there is enough time in the 
collision which passes through the deconfinement phase to establish 
near-chemical equilibrium conditions locally. 
The near chemical equilibrium leads to abundance anomalies 
such as the predicted~\cite{Raf82} $\bar\Lambda /{\bar p} > 1$, found 
in all A--A collisions~\cite{NA35pb}. There is a priori no opportunity
in terms of reaction mechanisms and time constants for a confined
HG state to reach conditions consistent with this result. 
This result is generally viewed to be a signature for the primordial 
QGP phase, as such  abundances are expected~\cite{Raf82,KMR86}  if the 
HG state of matter has as source the deconfined soup of quarks and 
gluons. 

We did not discuss in depth the entropy excess (see, however, our
earlier report\cite{Rio95}), which also are inconsistent with pure HG 
state, but are  well understood in terms of deconfinement.

Our view is that the observed hadronic particle effects, and 
specifically strangeness, provide a simple, consistent interpretation
of the heavy ion collision data within the hypothesis
that a novel type of deconfined hadronic 
matter is formed in both S- and Pb-induced 
reactions on heavy nuclei, but that quite different 
initial conditions are reached in these two cases. Moreover, the 
longitudinal flow, clearly visible in the S--S 200A GeV, in
{\it e.g.}, $\Lambda$-rapidity spectra, is not as pronounced 
in Pb--Pb reactions, but there seems to be relatively
strong transverse/radial flow driven by the high internal 
pressure which has been reached in the initially highly compressed 
fireball matter. We also expect that the much greater
volume of the fireball formed in Pb--Pb reactions, as compared to
 S--Pb leads to a greater volume of deconfined  initial phase, which 
in turn in the hadronization process should enhance the re-equilibration
of final state hadrons after freeze-out. Thus the Pb--Pb collisions 
should more approach the HG type chemical equilibrium of particle
yields than S--W/Pb or S--S reactions.

The only potentially contradiction to this unified view of the 
160--200 GeV A heavy ion reactions 
may be the systematics of charmonium production which 
suggest a difference in the absorption rate
of $c\bar c$ in central dense matter formed in Pb--Pb interactions as
compared to the earlier studies of pA and SA collisions. There is no
full and consistent theoretical understanding of all experimental 
details which may yet turn out to confirm our views 
about the presence of deconfinement in all these conditions.

\subsubsection{Immediate future}
Underlying the strange particle signatures of deconfinement 
is the rather rapid chemical equilibration of 
quark flavor, originating in gluon fusion reactions. 
Gluons themselves are the QGP fraction which is generally
believed to most rapidly approach chemical equilibrium in the
deconfined phase. We presumed in our studies that at SPS energies 
that gluon  chemical equilibrium has been reached, considering the 
lifespan of the reaction.  Such an assumption is less compelling 
at RHIC energies, indeed some model calculations exist arguing that
even gluons could not reach full equilibrium. These conclusions 
are reached assuming a fixed and small QCD interaction strength 
$\alpha_s$. Since there is no threshold to production of gluons,
and Bose effects enhance in dense matter gluon formation, our 
study of the strong scale dependence of $\alpha_s$ suggest
that one has to review the gluon chemical equilibration in the
near future. In anticipation  of this, we recommend use
of glue chemical equilibrium for RHIC energy scale collisions as well. 

The dynamics of evolving and exploding QGP phase impacts
significantly the resulting entropy distribution in rapidity and 
strange particle abundances. At present we have also engaged in 
several approaches to the explosive flow problem: we study 3-d 
hydrodynamic solutions,  develop semi-analytically soluble 
homologous hydrodynamic evolution/flow models of QGP and 
study the entropy sources during the flow phase. Our schematic 
approach~\cite{future} will permit by its simplicity 
a systematic investigation of the complex 
 dynamics of strongly interacting hadronic 
matter evolution comprises several interwoven
challenges:\\
a) the determination of initial flow conditions;\\
b) description of explosive flow of dense matter and the 
radiation reaction; \\
c) influence on matter flow by the equations of state \\
d) constrains on the final state conserved quantities 
(baryon number, entropy, and strangeness).\\
While hydrodynamical flow of thermally equilibrated matter 
is not entropy producing, other dynamical
 processes occur during the evolution 
of dense matter that  can produce entropy, for example
 chemical equilibration, i.e. approach of particle abundances to
their equilibrium number is entropy  producing. We consider of
 considerable importance that relativistic  hydrodynamical 
dynamical flow equations are improved 
to allow for entropy production by approach to phase  
space saturation by gluons and quarks.

The key problem we need to resolve is how to establish
that we have indeed seen the thermal, deconfined phase, 
where the degrees of freedom are those of nearly massless quarks and gluons?
Other forms and  phases of hadronic matter can  perhaps yet be invented 
that could produce similar  observational effects. 
We need to devise experimental approaches that would allow us to 
understand the rich deconfinement properties of  
dense relativistic nuclear/quark matter.
In immediate future we have to explore the  collision energy dependence
and see how the key properties change between the low AGS energies, 
the  intermediate SPS energies towards the immediate future of RHIC.

\vspace{0.5cm}
\subsection*{Acknowledgments}
I  would like to take this opportunity to thank  my
Brazilian   colleagues: \\
\indent {\bf Prof. Carlos E. Aguiar} of UFRJ; \\
\indent {\bf Dr. Sergio B. Duarte} of CBPF; \\
\indent {\bf Prof. Chung K. Cheong} of UERJ/CBPF; \\
\indent {\bf Prof. Yogiro Hama} of IF-USP; and the chairman, \\
\indent {\bf Prof. Takeshi Kodama}  of UFRJ \\
for putting together this 
most stimulating  and exciting week in Rio. 
I am sure that not just me, but we all,  have immensely enjoyed a rare 
combination of  the warm carioca spirit at the foot of the P\~{a}o 
de A\c{c}\'ucar,  with an intense, cutting edge, meeting program 
which brought us from far away together, kept us on the tip-toes
all day and left us  no time to breath. 

This report is dedicated 
to the many young participants at the workshop. 
Their enthusiasm was exemplary,
and their interest and commitment strengthened my resolve to 
compose a comprehensive write-up of my lecture. In doing so I 
have drawn on some material~\cite{acta97} developed originally
together with my long-time collaborators, and I would like to
thank in particular  J. Letessier and A. Tounsi from LPTHE-University 
Paris 7 for their kind support in our quest for strange particle 
signatures of QGP.

In the course of preparation of these notes I greatly benefited from 
interactions and comments made by Takeshi Kodama. I would like to thank
 him warmly for many valuable suggestions, and a careful reading of the 
manuscript draft.

\vfill\eject
This research program is supported
by US-Department of Energy under grant    
           DE-FG03-95ER40937\,.



\end{document}